\documentclass[usenatbib]{mn2e} 
\usepackage{graphicx,fixltx2e,amsmath,amssymb,amsfonts,latexsym}

\def\msun{\hbox{${\rm M}_{\odot}$}}
\def\mspy{\hbox{${\rm M}_{\odot}$\,yr$^{-1}$}}
\def\rsun{\hbox{${\rm R}_{\odot}$}}
\def\lsun{\hbox{${\rm L}_{\odot}$}}
\def\rcor{\hbox{$r_{\rm cor}$}}
\def\rmag{\hbox{$r_{\rm mag}$}}
\def\mstar{\hbox{$M_{\star}$}}
\def\rstar{\hbox{$R_{\star}$}}

\def\teff{\hbox{$T_{\rm eff}$}}
\def\logg{\hbox{$\log g$}}
\def\sn{\hbox{S/N}}
\def\vrad{\hbox{$v_{\rm rad}$}}

\def\kms{\hbox{km\,s$^{-1}$}}
\def\vsini{\hbox{$v \sin i$}}

\def\ptt{\hbox{$10^{-4} I_{\rm c}$}}

\def\degr{\hbox{$^\circ$}}

\def\Ip{\hbox{$I_{\rm p}$}}
\def\Vp{\hbox{$V_{\rm p}$}}
\def\Ipq{\hbox{$I_{\rm p,q}$}}
\def\Ipm{\hbox{$I_{\rm p,m}$}}
\def\Vpm{\hbox{$V_{\rm p,m}$}}

\def\Ie{\hbox{$I_{\rm e}$}}
\def\Ve{\hbox{$V_{\rm e}$}}
\def\Ik{\hbox{$I_{\rm k}$}}
\def\Ikq{\hbox{$I_{\rm k,q}$}}
\def\Ikm{\hbox{$I_{\rm k,m}$}}
\def\Ia{\hbox{$I_{\rm a}$}}
\def\Iaq{\hbox{$I_{\rm a,q}$}}
\def\Iam{\hbox{$I_{\rm a,m}$}}
\def\Vk{\hbox{$V_{\rm k}$}}
\def\Vkm{\hbox{$V_{\rm k,m}$}}
\def\Va{\hbox{$V_{\rm a}$}}
\def\Vam{\hbox{$V_{\rm a,m}$}}

\newcommand{\caii}{\hbox{Ca$\;${\sc ii}}}
\newcommand{\fei}{\hbox{Fe$\;${\sc i}}}

\newcommand{\hei}{\hbox{He$\;${\sc i}}}
\newcommand{\hal}{\hbox{H${\alpha}$}}
\newcommand{\hbe}{\hbox{H${\beta}$}}

\begin{document}

\title[Magnetospheric accretion and spin-down of AA~Tau]{Magnetospheric accretion and spin-down of the prototypical 
classical T~Tauri star AA~Tau } 

\makeatletter

\def\newauthor{%
  \end{author@tabular}\par
  \begin{author@tabular}[t]{@{}l@{}}}
\makeatother
 
\author[J.-F.~Donati et al.]
{\vspace{1.7mm}
J.-F.~Donati$^1$\thanks{E-mail: 
donati@ast.obs-mip.fr }, 
M.B.~Skelly$^1$, J.~Bouvier$^2$, S.G.~Gregory$^{3}$, K.N.~Grankin$^4$, \\ 
\vspace{1.7mm}
{\hspace{-1.5mm}\LARGE\rm 
M.M.~Jardine$^5$, G.A.J.~Hussain$^6$, F.~M\'enard$^2$, C.~Dougados$^2$, Y.~Unruh$^7$,} \\ 
\vspace{1.7mm}
{\hspace{-1.5mm}\LARGE\rm 
S.~Mohanty$^7$, M.~Auri\`ere$^1$, J.~Morin$^{8,1}$, R.~Far\`es$^1$ \& the MaPP collaboration} \\ 
$^1$ LATT--UMR 5572, CNRS \& Univ.\ de Toulouse, 14 Av.\ E.~Belin, F--31400 Toulouse, France \\
$^2$ LAOG--UMR 5571, CNRS \& Univ.\ J.~Fourier, 414 rue de la Piscine, F--38041 Grenoble, France \\ 
$^3$ School of Physics, Univ.\ of Exeter, Stocker Road, Exeter EX4~4QL, UK \\ 
$^4$ Crimean Astrophysical Observatory, Nauchny, Crimea 334413, Ukraine \\ 
$^5$ School of Physics and Astronomy, Univ.\ of St~Andrews, St~Andrews, Scotland KY16 9SS, UK \\
$^6$ ESO, Karl-Schwarzschild-Str.\ 2, D-85748 Garching, Germany \\ 
$^7$ Department of Physics, Imperial College London, London SW7 2AZ, UK \\ 
$^8$ Dublin Institute for Advanced Studies, School of Cosmic Physics, 31 Fitzwilliam Place, Dublin 2, Ireland 
}

\date{2010 June, MNRAS, submitted}
\maketitle
 
\begin{abstract}  

From observations collected with the ESPaDOnS spectropolarimeter at the 
Canada-France-Hawaii Telescope (CFHT) and with the NARVAL spectropolarimeter at the 
T\'elescope Bernard Lyot (TBL), we report the detection of Zeeman signatures on the 
prototypical classical T~Tauri star (cTTS) AA~Tau, both in photospheric lines and 
accretion-powered emission lines.  Using time series of unpolarized and circularly 
polarized spectra, we reconstruct at two epochs maps of the magnetic field, surface 
brightness and accretion-powered emission of AA~Tau.  We find that AA~Tau hosts a 
$2-3$~kG magnetic dipole tilted at $\simeq20$\degr\ to the rotation axis, and of 
presumably dynamo origin.  We also show that the magnetic poles of AA~Tau host large 
cool spots at photospheric level and accretion regions at chromospheric level.  

The accretion rate at the surface AA~Tau at the time of our observations (estimated 
from the emission in the \hei\ $D_3$ line mainly) is strongly variable, ranging from 
$-9.6$ to $-8.5$ and equal to $-9.2$ in average (in logarithmic scale and in \mspy);  
this is an order of magnitude smaller than the disc accretion rate at which the 
magnetic truncation radius (below which the disc is disrupted by the stellar magnetic 
field) matches the corotation radius (where the Keplerian period equals the stellar 
rotation period) -- a necessary condition for accretion to occur.  
It suggests that AA~Tau is largely in the propeller regime,  
with most of the accreting material in the inner disc regions being expelled outwards 
and only a small fraction accreted towards the surface of the star.  
The strong variability in the observed surface mass-accretion rate 
and the systematic time-lag of optical occultations (by the warped accretion disc) with 
respect to magnetic and accretion-powered emission maxima also support  
this conclusion.  

Our results imply that AA~Tau is being actively spun-down by the star-disc 
magnetic coupling and appears as an ideal laboratory for studying angular momentum 
losses of forming Suns in the propeller regime.  
\end{abstract}

\begin{keywords} 
stars: magnetic fields --  
stars: formation -- 
stars: imaging -- 
stars: rotation -- 
stars: individual:  AA~Tau --
techniques: spectropolarimetry 
\end{keywords}

\section{Introduction} 
\label{sec:int}

Classical T~Tauri stars (cTTSs) are young low-mass stars still contracting towards 
the main sequence and surrounded by gaseous and dusty accretion discs;  
they represent the important formation stage that stars with masses lower than 
$\simeq3$~\msun\ undergo at an age of a few Myrs, and during which they build up 
their exotic planetary systems.  

Observations reveal that magnetic fields play a crucial role at the cTTS stage.  
Thanks to strong large-scale fields \citep[e.g.,][for a review]{Donati09}, cTTSs 
are able to evacuate the core regions of their accretion discs, and to connect to 
the inner rim of their discs via discrete accretion funnels or veils through 
which material is accreted and angular momentum is dissipated \citep[e.g.,][for 
a review]{Bouvier07}.  Accretion discs are also expected to be magnetic, with 
fields enhancing accretion rates, generating powerful jets and modifying planet 
formation/migration mechanisms.  Surveys of the magnetic properties of cTTSs and 
their accretion discs are thus critically needed to understand the early history 
of low-mass stars in general and of the Sun in particular.  

Magnetic Protostars and Planets (MaPP) is an international project focussing 
specifically on this issue.  It has been granted 690~hr of observing time over 9 
consecutive semesters (2008b to 2012b) with the ESPaDOnS spectropolarimeter on 
the 3.8-m Canada-France-Hawaii Telescope (CFHT) to survey 15~cTTSs and 3 protostellar 
accretion discs of FU~Ori type (FUOrs);  it also regularly benefits from 
contemporaneous observations with the NARVAL spectropolarimeter on the 2-m T\'elescope 
Bernard Lyot (TBL) as well as photometric observations from Crimea, Uzbekistan and 
Armenia.  Additional multiwavelength observations from space (e.g., XMM-Newton/Chandra/HST) 
and/or from the ground (e.g., HARPS) are also organised in conjunction with MaPP 
campaigns on a few specific targets, providing deeper insights into the physical processes 
under scrutiny (and in particular magnetospheric accretion).  

MaPP primarily collects spectropolarimetric data probing the large-scale magnetic 
fields of cTTSs through the Zeeman signatures they generate in photospheric line 
profiles;  it also allows the detection of Zeeman signatures in the accretion spots 
located at the footpoints of accretion funnels.  By monitoring these Zeeman 
signatures over several successive rotation cycles (to filter out intrinsic 
variability and to retrieve rotational modulation more reliably and efficiently), 
MaPP can reconstruct maps of the large-scale magnetic fields of cTTSs and 
simultaneously recover the location of accretion spots.  By extrapolating 
from surface magnetograms, one can finally obtain an approximate 3D description of 
the magnetosphere, allowing more realistic and quantitative models of the geometry 
of accretion funnels and more generally of the star/disc magnetic coupling.  
Initial pre-MaPP studies of a few cTTSs were presented in several publications 
\citep{Donati07, Donati08, Jardine08, Gregory08, Hussain09, Donati10} to validate 
the main assumptions underlying the imaging code and demonstrate the overall feasibility 
of the modelling.  

The present paper is the first in a series dedicated to MaPP data and results,  
using a more general and better suited set of modelling assumptions and a more 
mature version of the imaging code;  ultimately, the main goal is to provide 
estimates and statistics on the topologies of large-scale magnetic fields of cTTSs 
and on the locations of their accretion spots, allowing the diagnosis of how they 
correlate with, e.g., stellar masses and rotation rates \citep[in a way similar 
to that achieved on main-sequence stars,][]{Donati09}, but also on parameters more 
specific to cTTSs, e.g., ages and accretion/ejection properties.  
This first MaPP study concentrates on the prototypical cTTS AA~Tau (see 
Sec.~\ref{sec:aatau} for a quick summary of the main stellar parameters relevant 
to this study).  We report here spectropolarimetric and photometric observations 
of AA~Tau (Sec.~\ref{sec:obs}), describe the variations of photospheric lines and 
accretion proxies (Sec.~\ref{sec:var}) and their subsequent modelling (Sec.~\ref{sec:mod}).  
We finally discuss the implications of these new results for our understanding of 
magnetospheric accretion processes in cTTSs (Sec.~\ref{sec:dis}).

\section{AA~Tau}
\label{sec:aatau}

AA~Tau is a well-known cTTS showing strong \hal\ emission and IR excesses 
demonstrating the presence of a gaseous and dusty accretion disc surrounding the protostar 
\citep[e.g.,][]{Bouvier99}.  A fit to the observed BVRI photometric fluxes 
\citep[showing $B-V$, $V-R_{\rm c}$ and $V-I_{\rm c}$ colors of 1.5, 0.9 and 1.9 
respectively at times of maximum brightness and minimum accretion, see Figs.~3 and 
6 of][]{Bouvier03} suggests that AA~Tau has a photospheric temperature of 
$\teff\simeq4000$~K and a visual reddening of $A_{\rm V}\simeq0.8$~mag \citep{Bouvier99}, 
implying a visual bolometric correction of $-1.8$ \citep{Bessell98}.  Given its distance 
($\simeq140$~pc) and maximum visual brightness \citep[corresponding to a magnitude of 
$\simeq$12.3, e.g.,][]{Grankin07}, AA~Tau has a bolometric luminosity equal to that of 
the Sun (within about 0.1~dex) and thus a radius of $\rstar\simeq2$~\rsun, suggesting a 
mass of $\mstar\simeq0.7$~\msun\ and an age of about 1.5~Myr \citep{Siess00} in agreement 
with previous estimates \citep[e.g.,][]{Bouvier99}.  This implies in particular that 
AA~Tau is fully convective and very similar (in mass, radius, age and rotation rate) to 
the other prototypical cTTS BP~Tau \citep{Donati08}.  

AA~Tau is known to undergo periodic eclipses with irregular depths, likely caused by 
partial occultations by a warped accretion disc viewed close to edge-on \citep{Bouvier99, 
Bouvier03, Bouvier07b}.  As the warp in the accretion disc is presumably caused by the 
magnetic field of the star, the recurrence period of these eclipses, equal to 
$8.22\pm0.03$~d \citep{Bouvier07b}, is actually tracing the rotation period of the star.  
Given that $\rstar\simeq2$~\rsun, the estimated rotation period implies that the equatorial 
rotation velocity of AA~Tau is 12.3~\kms;  the measured line-of-sight projected rotation 
velocity \citep[denoted \vsini\ where $i$ is the inclination of the rotation axis to the line 
of sight, and equal to $11.3\pm0.7$~\kms][]{Bouvier03} independently confirms that AA~Tau 
is viewed from the Earth at an inclination angle of about $70\pm10$\degr.  
With these parameters, the radius at which the Keplerian period is equal to 
the rotation period of the star, called the corotation radius and denoted \rcor, is equal 
to $\simeq7.6$~\rstar\ or 0.07~AU.  

Mass accretion on AA~Tau is reported to be smaller than the average rate expected for cTTSs of 
similar masses \citep[e.g.,][]{Johns07}.  The spectrum of AA~Tau exhibits all the usual 
accretion proxies, in particular \hal, \hbe, \hei\ $D_3$ and \caii\ infrared triplet 
(IRT) emission.  From their strength (in particular that of \hei\ emission) and the 
corresponding line fluxes (see Sec.~\ref{sec:obs}) in our spectra, and using empirical correlations 
from the published literature \citep[e.g.,][]{Fang09}, we can estimate the logarithmic mass-accretion 
rate at the surface of AA~Tau (in \mspy), found to vary from $-9.6$ to $-8.5$ throughout our runs 
(e.g., in 2008 December) and equal to about $-9.2$ in average.  
This is up to 10 times smaller than the accretion rate of BP~Tau 
\citep[equal to $-8.6$ when estimated with the same method,][]{Donati10}.  
Accretion on AA~Tau is known to be intrinsically variable on a time scale of a 
few days \citep[e.g.,][]{Bouvier03, Bouvier07b}.  
Optical veiling, i.e., the apparent weakening of the photospheric spectrum 
(presumably caused by accretion), is also often observed at a moderate (and variable) 
level on AA~Tau \citep[e.g.,][]{Bouvier03, Bouvier07b}.

\section{Observations}
\label{sec:obs}

Spectropolarimetric observations of AA~Tau were collected in 2008~December and 
2009~January using ESPaDOnS on the CFHT.  ESPaDOnS collects stellar spectra spanning 
the whole optical domain (from 370 to 1,000~nm) at a resolving power of 65,000 
(i.e., 4.6~\kms) and with a spectral sampling of 2.6~\kms, in either circular or 
linear polarisation \citep{Donati03}.  A total of 18 circular polarisation spectra 
were collected in 2 separate blocks shifted by about 1~month, with 11 spectra 
over a period of 15~d in 2008~December and 7 spectra over a period of 8~d in 
2009~January;  all polarisation spectra consist of 4 individual subexposures lasting 
each 1200~s and taken in different polarimeter configurations to allow the removal of 
all spurious polarisation signatures at first order.
Six additional spectra were collected a year before (in 2007~December and 2008~January) 
over a period of 12~d, with the ESPaDOnS twin NARVAL on the TBL, with slightly shorter 
exposure times.  

All raw frames are processed with {\sc Libre~ESpRIT}, a fully automatic reduction
package/pipeline available at CFHT and TBL.  It automatically performs optimal 
extraction of ESPaDOnS unpolarized (Stokes $I$) and circularly polarized (Stokes $V$) 
spectra grossly following the procedure described in \citet{Donati97b}.
The velocity step corresponding to CCD pixels is about 2.6~\kms;  however, thanks
to the fact that the spectrograph slit is tilted with respect to the CCD lines,
spectra corresponding to different CCD columns across each order feature a
different pixel sampling.  {\sc Libre~ESpRIT} uses this opportunity to carry out
optimal extraction of each spectrum on a sampling grid denser than the original
CCD sampling, with a spectral velocity step set to about 0.7 CCD pixel
(i.e.\ 1.8~\kms).
All spectra are automatically corrected of spectral shifts resulting from
instrumental effects (e.g., mechanical flexures, temperature or pressure variations) 
using telluric lines as a reference.  Though not perfect, this procedure provides 
spectra with a relative radial velocity (RV) precision of better than 0.030~\kms\
\citep[e.g.,][]{Donati08b}.

\begin{table}
\caption[]{Journal of ESPaDOnS/CFHT observations collected in 2008 December and 2009 
January.  Columns $1-4$ respectively list the UT date, the heliocentric Julian date and 
UT time (both at mid-exposure), and the peak signal to noise ratio (per 2.6~\kms\ 
velocity bin) of each observation (i.e., each sequence of $4\times1200$~s subexposures).  
Column 5 lists the rms noise level (relative to the unpolarized continuum level 
$I_{\rm c}$ and per 1.8~\kms\ velocity bin) in the circular polarization profile 
produced by Least-Squares Deconvolution (LSD), while column~6 indicates the 
rotational cycle associated with each exposure (using the ephemeris given by 
Eq.~\ref{eq:eph}).  }   
\begin{tabular}{cccccc}
\hline
Date & HJD          & UT      &  \sn\  & $\sigma_{\rm LSD}$ & Cycle \\
     & (2,454,000+) & (h:m:s) &      &   (\ptt)  &  (1+) \\
\hline
Dec 06 & 806.97106 & 11:11:06 & 180 & 2.5 & 48.510 \\
Dec 07 & 807.92124 & 09:59:24 & 200 & 2.3 & 48.625 \\
Dec 08 & 808.94040 & 10:27:02 & 160 & 3.0 & 48.749 \\
Dec 09 & 809.94201 & 10:29:24 & 130 & 3.7 & 48.871 \\
Dec 10 & 810.92881 & 10:10:27 & 100 & 5.8 & 48.991 \\
Dec 15 & 815.92556 & 10:06:02 & 190 & 2.4 & 49.599 \\
Dec 16 & 816.90946 & 09:42:55 &  60 & 8.7 & 49.719 \\
Dec 17 & 817.90170 & 09:31:49 & 140 & 3.7 & 49.840 \\
Dec 18 & 818.89873 & 09:27:36 & 160 & 3.2 & 49.961 \\
Dec 19 & 819.89824 & 09:26:57 & 110 & 4.6 & 50.082 \\
Dec 20 & 820.89756 & 09:26:02 & 120 & 4.3 & 50.204 \\
\hline
Jan 07 & 838.78706 & 06:48:28 & 160 & 2.8 & 52.380 \\
Jan 09 & 840.75479 & 06:02:13 & 180 & 2.6 & 52.620 \\
Jan 10 & 841.74685 & 05:50:53 & 190 & 2.3 & 52.740 \\
Jan 11 & 842.89978 & 09:31:14 & 140 & 3.8 & 52.881 \\
Jan 12 & 843.79311 & 06:57:43 & 110 & 4.9 & 52.989 \\
Jan 13 & 844.73006 & 05:27:02 & 130 & 3.9 & 53.103 \\
Jan 14 & 845.75784 & 06:07:09 & 180 & 2.7 & 53.228 \\
\hline
\end{tabular}
\label{tab:logesp}
\end{table}

\begin{table}
\caption[]{Same as Table~\ref{tab:logesp} for the additional NARVAL/TBL observations 
collected in 2007 December and 2008 January.  The exposure time of each sequence slightly 
varies (from $4\times1000$ to $4\times1200$~s) from night to night.  } 
\begin{tabular}{cccccc}
\hline
Date & HJD          & UT      &  \sn\  & $\sigma_{\rm LSD}$ & Cycle \\
     & (2,454,000+) & (h:m:s) &      &   (\ptt)  &  (1+) \\
\hline
Dec 27 & 462.36191 & 20:35:13 & 70 & 10.0 & 6.587 \\
Dec 31 & 465.54716 & 01:02:15 & 80 &  7.0 & 6.974 \\
Dec 31 & 466.37216 & 20:50:19 & 90 &  6.1 & 7.075 \\
Jan 01 & 467.37235 & 20:50:41 & 80 &  7.7 & 7.196 \\
Jan 03 & 468.55236 & 01:10:01 & 60 &  8.3 & 7.340 \\
Jan 07 & 473.38037 & 21:02:49 & 70 &  7.2 & 7.927 \\
\hline
\end{tabular}
\label{tab:lognar}
\end{table}

The peak signal-to-noise ratios (\sn, per 2.6~\kms\ velocity bin) achieved on the
collected spectra (i.e., the sequence of 4 subexposures) range between 100 and
200 for ESPaDOnS data (except for one spectrum recorded in poor weather conditions) 
and between 60 and 90 for NARVAL data (directly reflecting the smaller collecting 
area).  Rotational cycles $E$ are computed from heliocentric Julian dates 
according to the ephemeris:  
\begin{equation}
\mbox{HJD} = 2454400.0 + 8.22 E 
\label{eq:eph}
\end{equation}
where the rotation period is taken from \citet{Bouvier07b}.  
The full journal of observations is presented in Tables~\ref{tab:logesp} \& \ref{tab:lognar} 
for ESPaDOnS/CFHT and NARVAL/TBL data respectively.

Least-Squares Deconvolution \citep[LSD,][]{Donati97b} was applied to all
observations.   The line list we employed for LSD is computed from an {\sc
Atlas9} LTE model atmosphere \citep{Kurucz93} and corresponds to a K7 
spectral type ($\teff=4,000$~K and  $\logg=3.5$) appropriate for AA~Tau.
Only moderate to strong atomic spectral lines (with line-to-continuum core 
depressions larger than 40\% prior to all non-thermal broadening) are included 
in this list;  spectral regions with strong lines mostly formed outside the 
photosphere (e.g., Balmer, He, \caii\ H, K and IRT lines) and/or heavily crowded 
with telluric lines were discarded.  
Altogether, more than 9,000 spectral features are used in this process, with about 
40\% of them from \fei.  Expressed in units of the unpolarized continuum level
$I_{\rm c}$, the average noise levels of the resulting Stokes $V$ LSD signatures 
are ranging from 2.3 to 5.7$\times10^{-4}$ per 1.8~\kms\ velocity bin for ESPaDOnS 
data (except on Dec~16) and from 6 to 10$\times10^{-4}$ for NARVAL data.  

For estimating when eclipses of AA~Tau are occurring, contemporaneous photometry was 
collected from Crimean Astrophysical Observatory (CrAO) over a period extending about 1~month 
before and after the main spectropolarimetric runs;  a total of 5 and 13 measurements 
were obtained in conjunction with our 2008/2009 and 2007/2008 runs respectively.  
Additional photometric data from the All-Sky Automated Survey \citep[ASAS,][]{Pojmanski97} 
were added to improve phase coverage;  selecting only grade A exposures, and further 
rejecting statistically deviant points, we are left with 12 and 9 supplementary ASAS 
measurements contemporaneous with our 2008/2009 and 2007/2008 runs.  
Typical rms photometric accuracies 
are about 50 and 100~mmag for CrAO and ASAS data respectively.

\section{Variations of photospheric lines and accretion proxies}
\label{sec:var}

Zeeman signatures with typical peak-to-peak amplitudes of about 1\% are clearly detected 
at all times in LSD profiles of photospheric lines (see Fig.~\ref{fig:pol} top right panel);  
temporal variations are also detected but remain moderate, the average Stokes $V$ profile over the 
whole run featuring a roughly symmetric shape (with respect to line centre) suggesting the 
presence of a significant toroidal field component at the surface of the star 
\citep[e.g.,][]{Donati05}.  
As a result, the line-of-sight projected component of the field averaged over 
the visible stellar hemisphere \citep[called longitudinal field and estimated from the first 
moment of the Stokes $V$ profile,][]{Donati97b} is weak, varying from $-230$~G to $+70$~G 
during the 2008/2009 observing run depending on the epoch (see Fig.~\ref{fig:var} lower left 
panel).  

LSD Stokes $I$ profiles of photospheric lines also vary with time, both in position and 
strength (see Fig.~\ref{fig:pol} top left panel).  The corresponding RV variations reach 
a peak-to-peak amplitude of about 2~\kms\ (about a mean of about 17~\kms), similar to 
previous results reporting that these variations are apparently stable on a long-term basis 
and tentatively attributed to spots at the surface of AA~Tau \citep{Bouvier07b}.  
By comparing our spectra of AA~Tau to those 
of a spectroscopic template of similar spectral type, we can in principle retrieve estimates 
of the veiling, i.e., the amount by which photospheric lines are weakened with respect to 
those of the template star.  The template star we observed (61~Cyg~B, of similar temperature 
but different luminosity class) shows LSD profiles that are weaker (by $10-30$\%) than most of our 
AA~Tau profiles.  While it first indicates that a better template star is needed to obtain 
absolute veiling estimates, it at least suggests that veiling is only moderate for AA~Tau 
at the time of our observations.  Using our strongest LSD profile of AA~Tau (at cycle 
53.103) as a zero-veiling reference, we find that veiling is most of the time smaller than 
5\% during the 2008/2009 observing run, except between rotation cycles $49.7-50.2$ where it 
peaks at about 20\% (see Fig.~\ref{fig:var} lower right panel).  

We find that LSD Stokes $I$ and $V$ profiles collected at similar rotation phases can 
exhibit different shapes at different rotation cycles (e.g., at cycles 48.991, 49.961 and 
52.989), indicating that intrinsic variability is significant.  In particular, veiling 
does not repeat well between successive rotation cycles (see Fig.~\ref{fig:var} lower 
right panel), suggesting that this intrinsic variability likely reflects unsteady accretion 
at the surface of AA~Tau.  We note that veiling remains smaller than 5\% throughout our 
2009 January observations (rotation cycle $52.3-53.3$), suggesting that we are catching 
AA~Tau in a state of constantly very low accretion at this time of the 2008/2009 run;  
we suspect that, as a result, this is also an epoch at which rotational modulation of 
spectral lines is easiest to detect against intrinsic variability.  
We finally note that strong veiling (when present) 
occurs at rotation phases $0.8-1.1$ in the 2008/2009 run, i.e., slightly before eclipse 
maximum (at phase 1.05, see Fig.~\ref{fig:var} upper panel) and similar to previous 
reports \citep[e.g.,][]{Bouvier07b}.  

Usually considered as the most reliable accretion proxy, \hei\ $D_3$ emission at 587.562~nm 
is thought to be produced in the postshock region at the footpoints of accretion funnels, 
i.e., in chromospheric accretion spots.  At the time of our observations, \hei\ emission 
always shows up as a narrow profile (see Fig.~\ref{fig:bal}, top panel) and 
amounts to an average equivalent width of about 15~\kms\ (i.e., 0.030~nm).  It is strongly 
variable with time, with equivalent widths varying from 7.5 to 45~\kms\ (i.e., 0.015 to 
0.090~nm) throughout our 2008/2009 run (see Fig.~\ref{fig:var} second panel right column).  
Previous studies \citep[e.g.,][]{Bouvier03, Bouvier07b} report very similar levels of \hei\ 
emission, equal to about 25~\kms\ in average (i.e., 0.05~nm), most of the time lower than 
50~\kms\ (i.e., 0.10~nm) and reaching at most 75~\kms\ (i.e., 0.15~nm) during highest-accretion 
episodes.  
Times of strongest \hei\ emission clearly coincide with epochs of maximum veiling, as 
previously reported in the literature \citep[e.g.,][]{Bouvier07b};  in particular, using the 
published correlation between veiling and \hei\ emission \citep[see Fig.~5 of][]{Bouvier07b}, 
we can safely confirm that veiling is indeed very small at rotational cycle 53.103 (assumed 
as our zero-veiling reference epoch, see above).  Most of the observed variation of \hei\ 
emission does not repeat from one rotation cycle to the next and thus mostly reflects 
intrinsic variability rather than rotational modulation.  We find that \hei\ emission is 
minimal throughout our 2009 January observations, with equivalent widths varying from 7.5 to 
10.2~\kms\ (i.e., 0.015-0.020~nm);  it confirms that rotation cycle $52.3-53.3$ corresponds to 
a very-low-accretion stage of AA~Tau with limited (i.e., $\simeq$30\% peak-to-peak) though 
definite rotational modulation, and with maximum emission occurring around phase 1.0.  

Clear Zeeman signatures are detected in conjunction with \hei\ $D_3$ emission, corresponding 
to longitudinal fields as strong as $-2.5$~kG (see Fig.~\ref{fig:var} second panel left 
column) and similar to previous published reports \citep[e.g.,][]{Valenti04}.  We note 
that longitudinal fields are not markedly different in phases of low and high accretion, 
indirectly confirming that \hei\ emission is mostly probing accretion spots at the footpoints 
of accretion funnels;  as a result, intrinsic variability on this longitudinal field 
estimate is limited while rotational modulation is fairly clear.  Longitudinal field is 
strongest around phase 1.0 (i.e., in phase with line emission), indicating that this is 
when the accretion spot is best visible to an Earth-based observer.  

\begin{figure*}
\vspace{-2mm}
\center{\hbox{\includegraphics[scale=0.6,angle=-90]{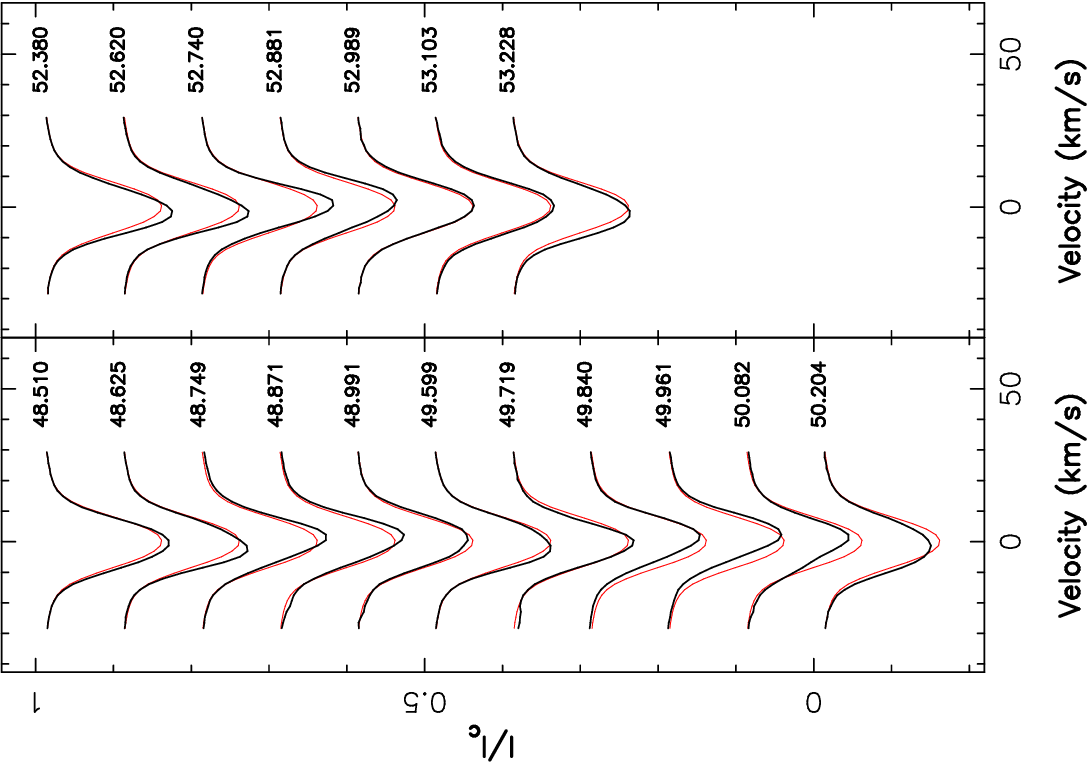}\hspace{4mm}
\includegraphics[scale=0.6,angle=-90]{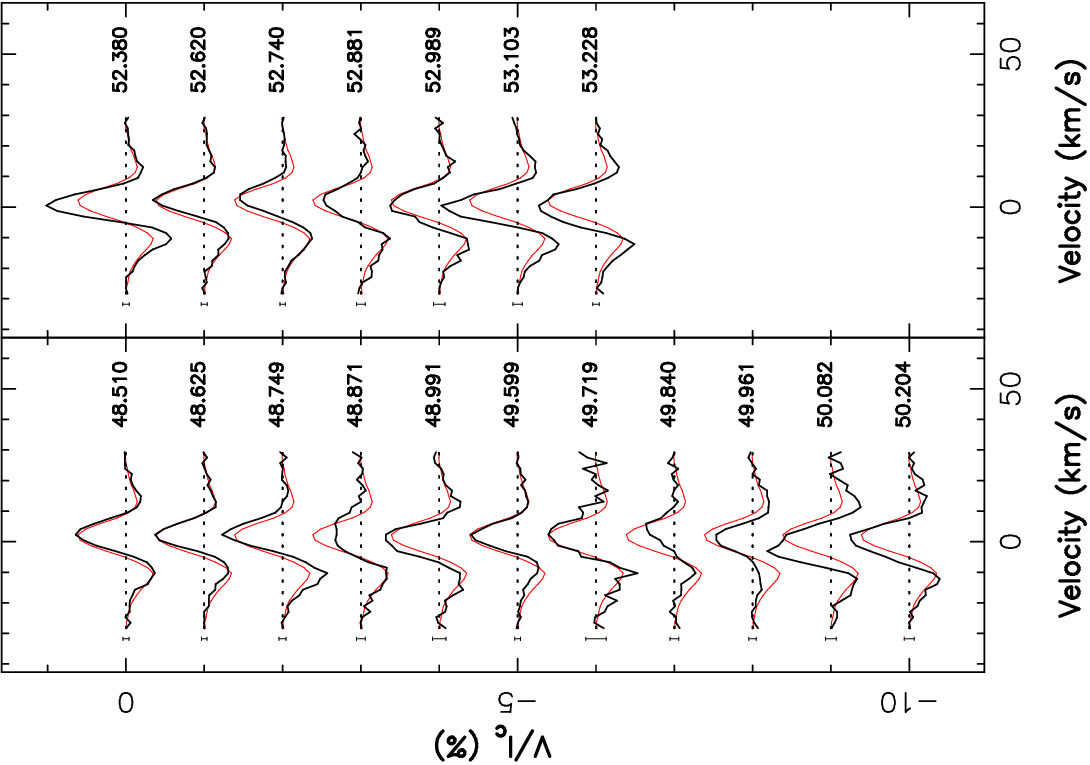}}} 
\vspace{5mm}
\center{\hbox{\includegraphics[scale=0.6,angle=-90]{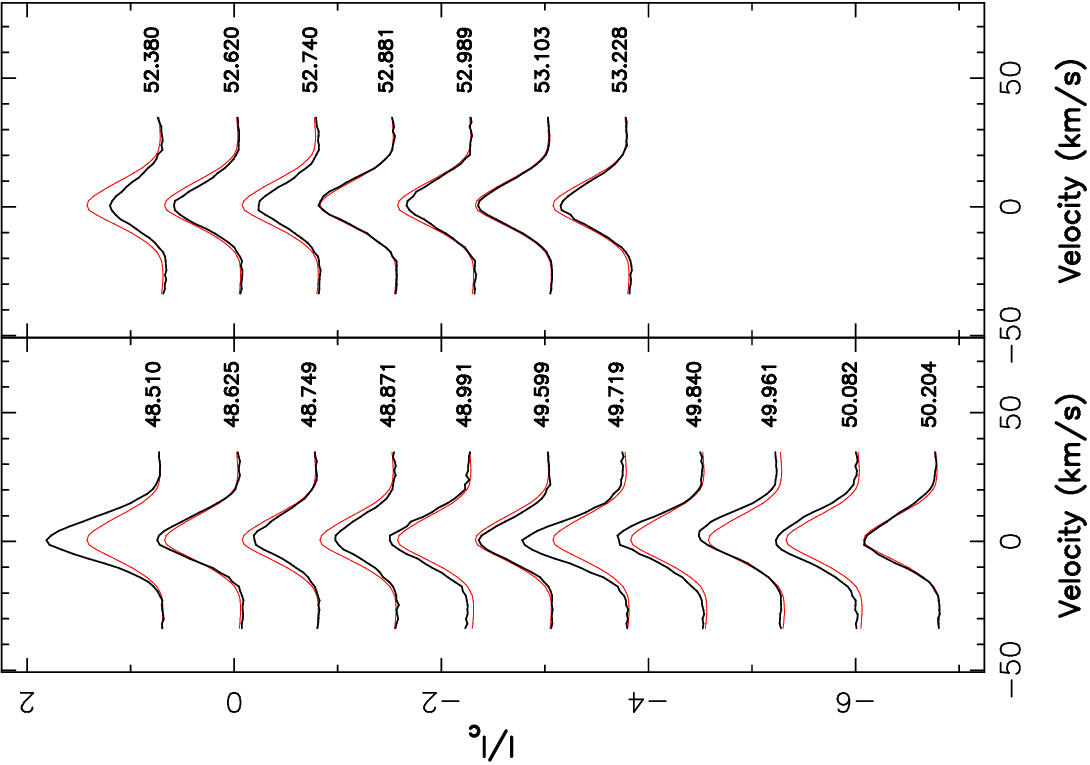}\hspace{4mm}
\includegraphics[scale=0.6,angle=-90]{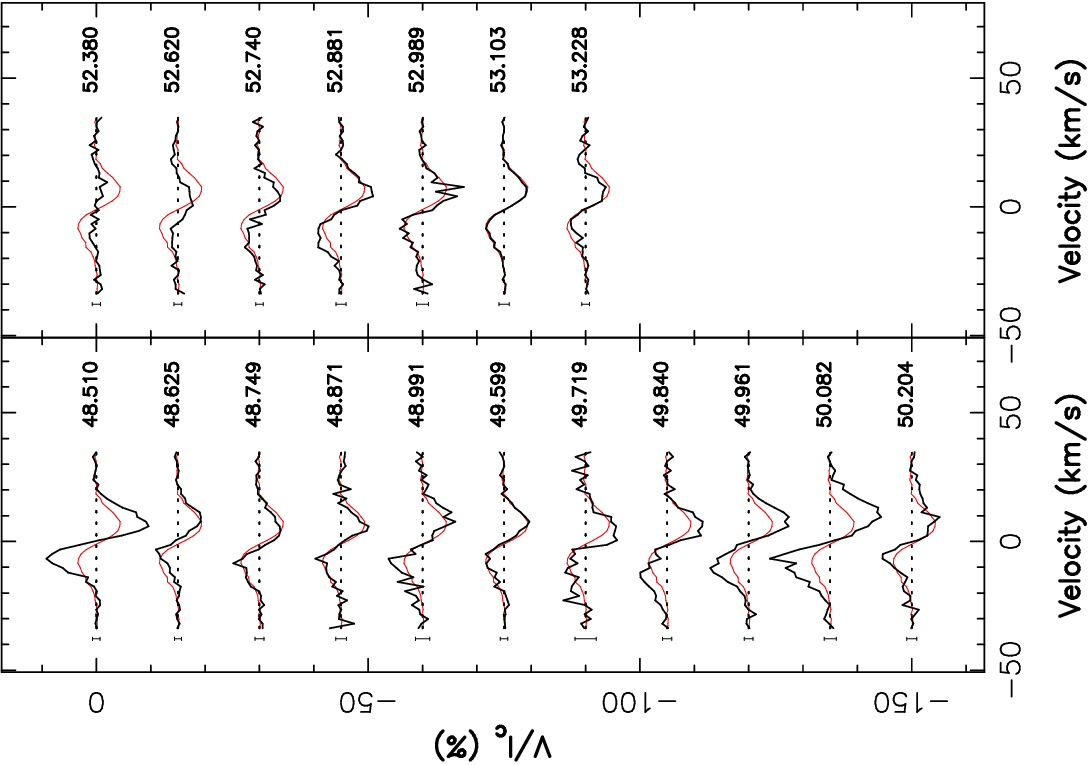}}} 
\caption[]{Temporal variations of the Stokes $I$ (left) and Stokes $V$ (right) LSD profiles 
of the photospheric lines (top) and of the \caii\ emission 
(averaged over the 3 IRT lines, 
bottom) of AA~Tau in 2008 December (left column of each panel) and 2009 January (right column).  
The Stokes $I$ profiles of \caii\ emission is shown before subtracting the underlying (much 
wider) photospheric absorption, hence the reduced flux in the far wings (with respect to the 
unit continuum).  
To emphasize variability, the average profile over the run is shown in red.  Rotation cycles 
(as listed in Table~1) and 3$\sigma$ error bars (for Stokes $V$ data only) are also included 
next to each profile.  } 
\label{fig:pol}
\end{figure*}

\begin{figure*}
\center{\hbox{\includegraphics[scale=1.0,angle=-90]{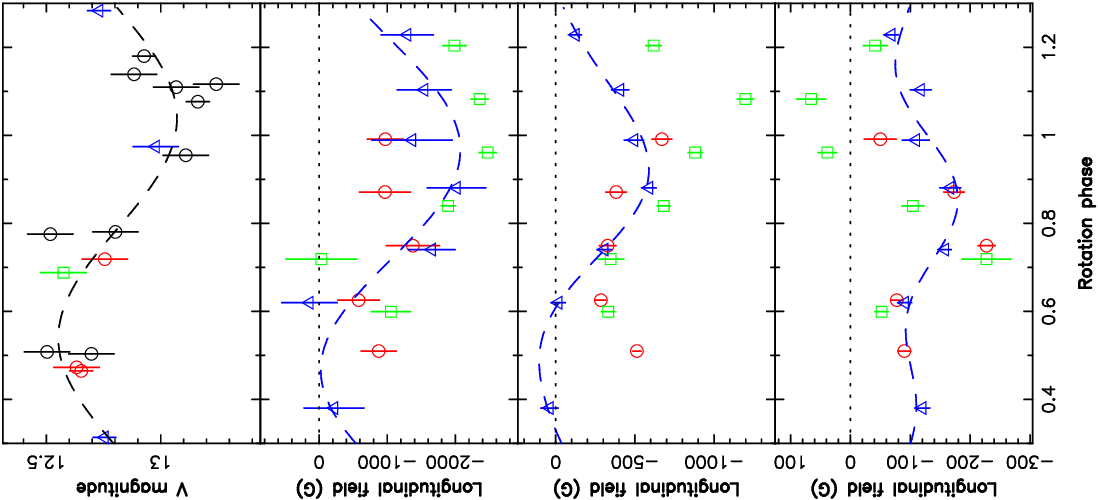}\hspace{4mm}
\includegraphics[scale=1.0,angle=-90]{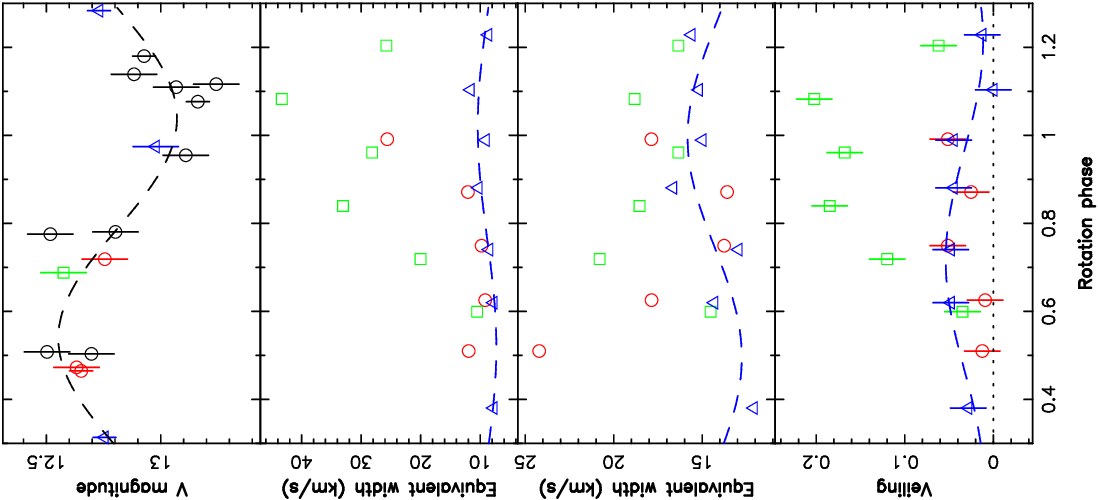}}} 
\caption[]{Temporal variations of the integrated brightness (top panels), \hei\ $D_3$ (second 
panels), \caii\ IRT (third panels) and photospheric lines (bottom panels) of AA~Tau in 2008 
December and 2009 January.  
Longitudinal field variations are shown on the left side (3 lower panels) while equivalent 
width and veiling variations are shown on the right side.  Data 
collected within cycles $48.3-49.3$, $49.3-50.3$ and $52.3-53.3$ are respectively plotted 
as red circles, green squares and blue triangles.  Fits with sine/cosine waves (plus first 
overtones for the 2 lower left panels) are included (and shown as dashed lines) for data 
collected within rotation cycle $52.3-53.3$ (to outline the amount of variability caused 
by rotational modulation) and for all photometric data;  
$\pm1$~$\sigma$ error bars on data points are also shown whenever larger than symbols.  } 
\label{fig:var}
\end{figure*}

\begin{figure*}
\center{\hbox{\includegraphics[scale=0.6,angle=-90]{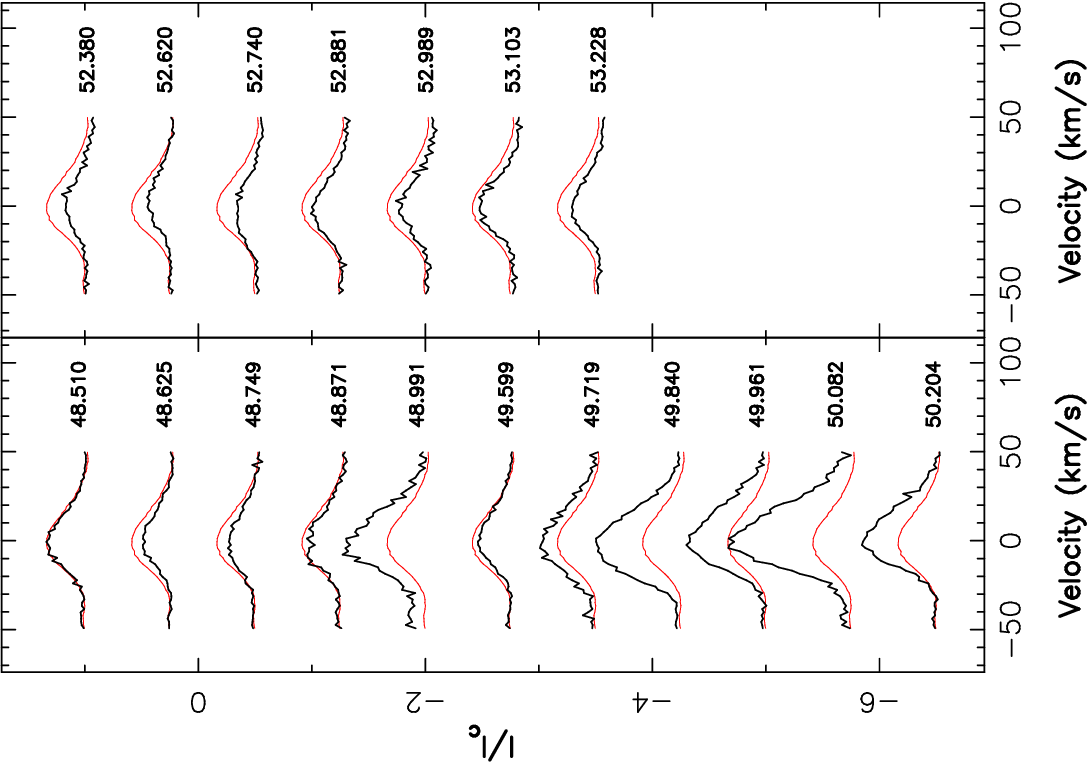}\hspace{4mm}
\includegraphics[scale=0.6,angle=-90]{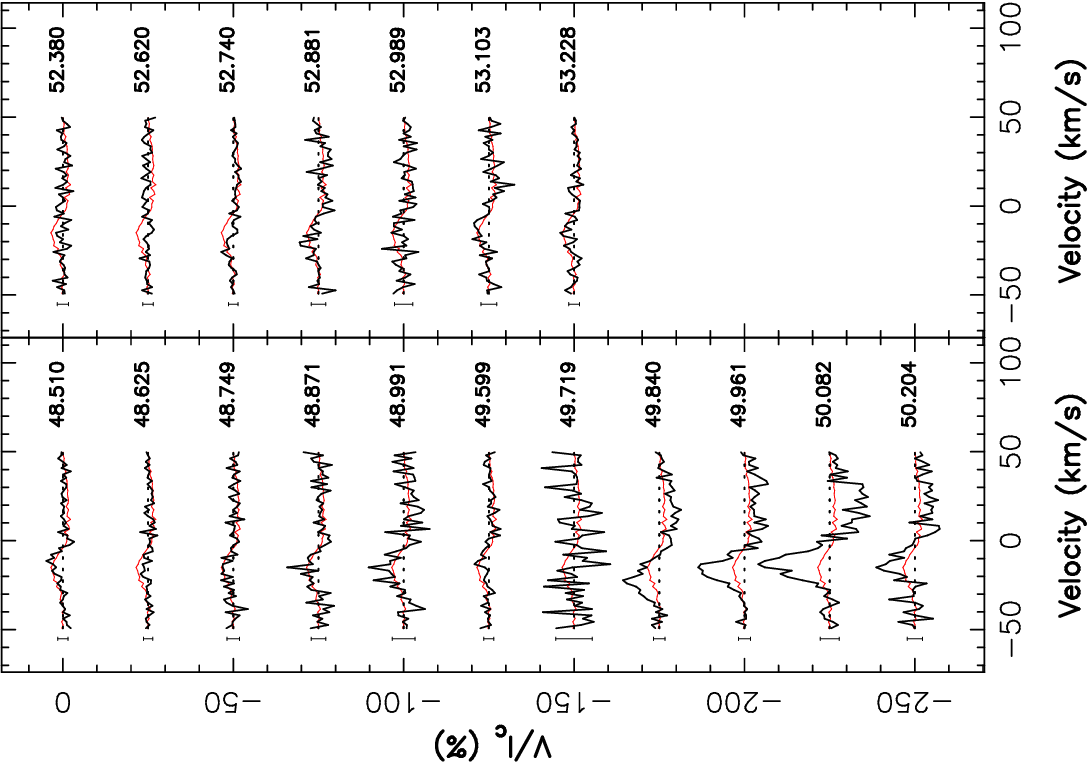}}} 
\vspace{5mm}
\center{\hbox{\includegraphics[scale=0.6,angle=-90]{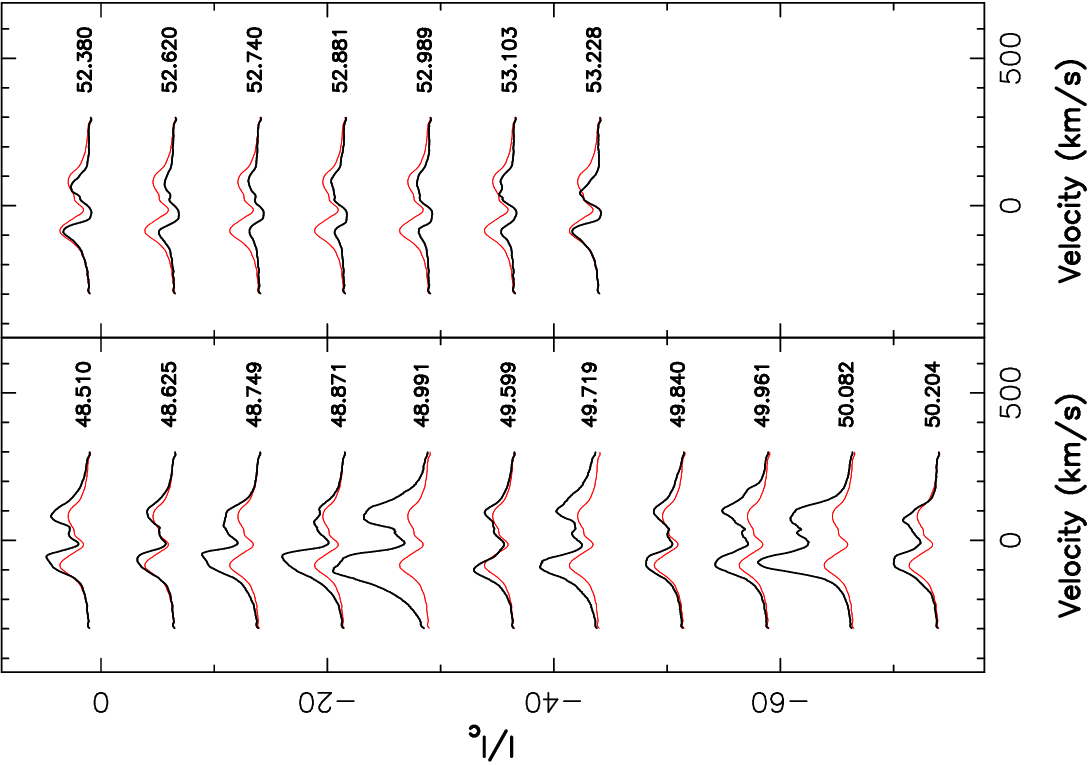}\hspace{4mm}
\includegraphics[scale=0.6,angle=-90]{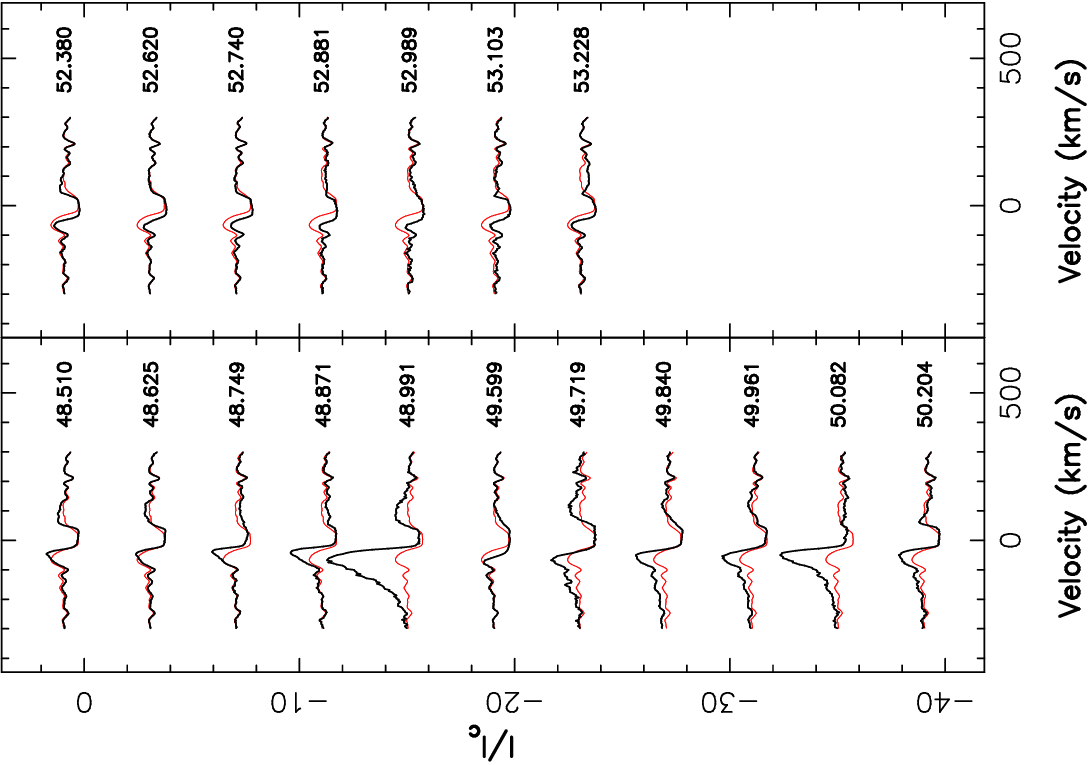}}} 
\caption[]{Variation of the \hei\ $D_3$ (Stokes $I$: top left, Stokes $V$: top right), 
H$\alpha$ (bottom left) and H$\beta$ (bottom right) lines of AA~Tau in 2008 
December (left column of each panel) and 2009 January (right column).  
Line fluxes in the far wings of all Stokes $I$ profiles are close to the (unit) continuum level.  
To emphasize variability, 
the average profile over the run is shown in red.  Rotation cycles (as listed in Table~1) and 
3$\sigma$ error bars (for Stokes $V$ data only) are also mentioned next to each profile.  } 
\label{fig:bal}
\end{figure*}

Core emission in \caii\ IRT lines are also useful for probing magnetospheric accretion.   
Since \caii\ emission is presumably also coming from the non-accreting chromosphere, it 
is less specifically related to accretion spots than \hei\ emission and is thus a more 
ambiguous proxy.  However, the redder spectral location, higher magnetic sensitivity and 
multiple nature of the corresponding spectral lines more than compensate for this drawback;  
moreover, the shape of the corresponding Zeeman signatures is simpler \citep[i.e., 
featuring a classical, nearly-antisymmetric pattern, as opposed to the strongly 
non-antisymmetric Zeeman signatures of the \hei\ line, e.g.,][]{Donati07, Donati08} 
and thus easier to model.  
To extract the core \caii\ emission profiles (used later in the imaging analysis, see 
Sec.~\ref{sec:mod}) from our spectra, we start by constructing a LSD-like weighted 
average of the 3 IRT lines;  we then subtract the underlying (much wider) photospheric 
absorption profile, with a Lorentzian fit to the far wings over a velocity interval of 
$\pm200$~\kms\ about the emission core.  
Equivalent widths of the \caii\ emission are in average 
equal to about 15~\kms\ (i.e., 0.043~nm) and vary from 12 to 24~\kms\ (i.e., 
$0.035-0.070$~nm) during our 2008/2009 run.  As for \hei,  most of the variation does 
not repeat between successive rotation cycles (see Fig.~\ref{fig:pol} bottom left panel 
and Fig.~\ref{fig:var} third panel right column) and is thus intrinsic in nature.  
Note however that the strongest \caii\ emission episode (at cycle 48.510) coincides 
with no more than a small increase in \hei\ emission (see Fig.~\ref{fig:var} right 
column) while large \hei\ emission episodes generate only moderate \caii\ emission;  
this directly relates to the fact that \caii\ emission is more sensitive to 
chromospheric activity and flares (the likely cause of the emission episode at cycle 
48.510) than \hei\ emission.  During rotation cycle $52.3-53.3$ (i.e., in 2009 January), 
\caii\ emission is close to minimum, varying by about 20\% peak to peak;  maximum 
emission is reached at about phase 1.0 (as for \hei\ emission, see Fig.~\ref{fig:var} 
middle panels right column), confirming that most of the observed rotational modulation 
at this time of the run is due to the chromospheric accretion spot.  

Zeeman signatures in \caii\ IRT lines are clearly detected at almost all epochs, with 
average peak-to-peak amplitudes of about 10\% (see Fig.~\ref{fig:pol} bottom right panel).  
Corresponding longitudinal fields reach up to about $-1$~kG in high accretion states ($-0.9$ 
and $-1.2$~kG at cycles 49.961 and 50.082 respectively) and more typically range from 0 to 
$-700$~G otherwise.  In 2009 January in particular (rotation cycles $52.3-53.3$), longitudinal 
fields vary very smoothly, and in phase with longitudinal fields derived from the \hei\ line;  
this is further evidence that the observed fluctuation mainly reflects rotational modulation.  
Maximum field is reached at phase 0.95, i.e., when the accretion spot faces the observer;  
maximum field strength is about 3 times weaker than that from \hei\ lines, suggesting that 
the accretion spot contributes no more than about one fourth to \caii\ emission (the 
remaining part coming from the non-accreting chromosphere).  

H$\alpha$ and H$\beta$ both feature emission components of varying strengths (see 
Fig.~\ref{fig:bal} bottom panel);  while H$\alpha$ emission is strong and wide and truncated by a 
central absorption component (giving the profile a mostly double-peak appearance), 
H$\beta$ emission is much weaker and mostly confined to the blue profile wing.  
During the 2008/2009 run, the average equivalent width of H$\alpha$ is 550~\kms\ 
(1.2~nm), with peaks of up to 2050~\kms\ (4.5~nm, at cycle 48.991).  For H$\beta$, the 
average equivalent width is close to 0 (within about 20~\kms\ or 0.03~nm), with the 
weak blue-wing emission barely compensating for the central absorption;  at cycle 
48.991, H$\beta$ emission peaks at an equivalent width of 430~\kms\ (0.7~nm).  

In 2009 January, H$\alpha$ is weakest, with an average equivalent width of 250~\kms\ 
(0.55~nm);  blue and red emission components vary more or less in phase with 
each other, peaking around phase 0.35 and minimum at phase 0.85.  
At the same time, H$\beta$ exhibits strong central absorption and reduced blue-wing 
emission, with an average equivalent width as low as $-70$~\kms\ ($-0.1$~nm) and 
profile variation concentrating mainly in the red wing (at about +100~\kms);  maximum 
and minimum emission occur at phase 0.5 and 1.0 respectively.  
Following \citet{Bouvier07b}, we speculate that the absorption episodes in the 
red wing emission of H$\beta$ trace accretion funnels crossing the line of sight as they 
corotate around the star.  

Using the average equivalents widths of the \hei\ line, the \caii\ IRT and H$\alpha$ 
during our 2008/2009 run, we derive logarithmic line fluxes (with respect to the luminosity 
of the Sun \lsun) equal to $-5.3$, $-5.1$ and $-3.6$ respectively.  This implies a logarithmic 
accretion luminosity (with respect to \lsun) of $-2.3\pm0.3$ using the empirical correlations 
of \citet{Fang09} and putting strong weight on the \hei\ data, and thus a logarithmic 
mass-accretion rate at the surface of the star of $-9.2\pm0.3$ (in \mspy).  
Mass accretion rates can also be estimated (though more roughly) through the full width 
of H$\alpha$ at 10\% height \citep[e.g.,][]{Natta04, cieza10};  in our case, the full 
widths we determine range from 350~\kms\ (in 2009 January) up to 480~\kms\ (at cycle 
48.991), with an average of about 420~\kms, yielding logarithmic mass-accretion rates 
of $-9.5$, $-8.8$ and $-8.2$ (with typical errors of $\pm0.6$) respectively 
supporting our main estimates.  
We find that the logarithmic surface mass-accretion rate at magnetic maximum (i.e., around phase 0.0) can 
vary by about an order of magnitude (from $-9.4$ to $-8.5$) from one cycle to the next, while 
that at magnetic minimum (i.e., around phase 0.5) is more stable on the long term (at about $-9.6$).  
During the low-accretion phase of 2009 January, the logarithmic mass-accretion rate at the 
surface of AA~Tau is found to be $-9.5$ in average and to vary by only $\simeq50$\% 
peak-to-peak (from $-9.6$ to $-9.4$).  

For our 2007/2008 run, we obtain similar results (not shown here) but with 
larger error bars and worse phase sampling (7 phases only, the first 6 roughly covering 
one cycle, see Table~\ref{tab:lognar}).  
Zeeman signatures in Stokes $V$ LSD profiles of photospheric lines are again detected 
at all phases, tracing longitudinal fields of $+90$ to $-220$~G.  
Emission in \caii\ and \hei\ lines vary smoothly with time, reaching maximum strengths 
at phase 1.3;  longitudinal fields from the \caii\ lines also show smooth changes 
from 0 to $-730$~G, reaching maximum field at phase 1.2.  This suggests in particular 
that the accretion spot is located at phase 1.2-1.3, i.e., again slightly before eclipse 
maximum (i.e., brightness minimum, occurring at phase 1.35).  
The variability observed in all spectral proxies is apparently compatible with 
rotational modulation;  the moderate phase coverage and poor phase redundancy achieved 
in this initial run does however not allow us to quantify reliably the relative strength 
of intrinsic variability.  
Emission in \hei, \caii\ IRT and H$\alpha$ lines feature equivalent widths of 14~\kms\ 
(0.027~nm), 14~\kms\ (0.040~nm) and 700~\kms\ (1.5~nm), corresponding to an average 
logarithmic mass-accretion rate of $-9.3\pm0.3$ (in \mspy) similar to that during the 2008/2009 run.  
The logarithmic surface mass-accretion rates at magnetic minimum and maximum are respectively equal 
to $-9.6$ and $-9.0$, i.e., showing a larger peak-to-peak variation (but a similar level at magnetic 
minimum) than in 2009 January.

\section{Modelling the surface of AA~Tau}
\label{sec:mod}

The visual inspection of all spectral proxies and their temporal variations (presented in 
Sec.~\ref{sec:var}) straightforwardly reveals a number of interesting results, in particular 
concerning the chromospheric spot at the footpoint of the accretion funnels linking AA~Tau with 
its accretion disc.  Assuming that the field within this accretion spot is mainly radial 
and generates the periodic longitudinal field fluctuation seen in the \hei\ line, we can 
derive from the observed rotational modulation that the accretion spot is located at a latitude 
of about 70\degr\ (to produce a null field extremum at phase 0.5 and given that $i\simeq70\degr$, 
see Sec.~\ref{sec:aatau}) and thus that the field strength within the spot is of order of 3~kG 
(generating longitudinal field maxima of about 2~kG when the field reaches its minimum inclination 
of 50\degr\ to the line of sight).  
Eclipse maxima are obviously linked with (though slightly lagging) longitudinal field maxima and 
peak absorption in the red wing of Balmer lines, confirming that eclipses are likely due 
to a disc warp at the base of the accretion veil, as initially proposed by \citet{Bouvier03} and 
further documented by \citet{Bouvier07b}.  

Going further, and in particular quantifying how magnetic fields are distributed at the surface 
of AA~Tau and investigating how fields connect the protostar to its accretion 
disc, requires a more detailed modelling of the observed profiles and their rotational modulation.  
This is what we propose below where we introduce a more mature and elaborate model that builds upon 
the preliminary version introduced in previous papers \citep{Donati07, Donati08}.  
We then apply the model to our spectra of AA~Tau collected in 2009 January (those from 2008 
December containing too much intrinsic variability, see Sec.~\ref{sec:var}) and in 2007/2008.  

\subsection{Mapping magnetic field, photospheric brightness and chromospheric accretion}

Our model is designed to recover simultaneously maps of the surface magnetic field, of the photospheric 
brightness and of the accretion-powered excess emission from sets of Stokes $I$ and $V$ LSD and \caii\ 
profiles of classical T~Tauri stars such as those shown in Fig.~\ref{fig:pol}.  
We do not attempt at matching \hei\ profiles in this paper, as the complex shape of their Zeeman 
signatures (see Sec.~\ref{sec:var}) would require some modelling of the velocity flows within the 
postshock region of accretion funnels, beyond the scope of our study. 
Similarly, a quantitative modelling of Balmer line profiles, requiring a full 3D description of the 
magnetosphere (including density and velocity fields), is postponed for future papers.  

In this model, the magnetic field is described through its poloidal and toroidal components expressed 
as spherical-harmonics (SH) expansions \citep{Donati06b}.  The spatial distribution of photospheric 
brightness (with respect to the quiet photosphere) and that of \caii\ excess emission (with respect 
to the quiet chromosphere) are modelled as series of independent pixels (typically a few thousand) 
on a grid covering the visible surface of the star;  brightness spots are assumed to be darker than 
the photosphere while accretion spots are supposed to be brighter than the chromosphere.  
The main difference with respect to the preliminary version of the code is that the photospheric 
and chromospheric images are no longer assumed to be homothetic but are fitted to the data
independently from one another;  moreover, the quantity that we now recover for the photospheric 
image is brightness (rather than spottedness).  

Following the principles of maximum entropy, the code automatically retrieves the simplest magnetic 
topology, brightness image and accretion map compatible with the series of rotationally modulated 
Stokes $I$ and $V$ LSD profiles.  
The reconstruction process is iterative and proceeds by comparing at each step the synthetic
Stokes $I$ and $V$ profiles corresponding to the current images with those of the observed data set.  
To compute the synthetic profiles, we add up the elementary spectral contributions from each image 
pixel over the visible hemisphere, taking into account all relevant local parameters of the 
corresponding grid cell (e.g., brightness, chromospheric excess emission, magnetic field strength 
and orientation, radial velocity, limb angle, projected area).  Since the problem is partly ill-posed, 
we stabilise the inversion process by using an entropy criterion (applied to the SH  
coefficients and to the brightness/excess emission image pixels) aimed at selecting the field 
topologies and images with minimum information among all those compatible with the data.  
The algorithm used to solve the maximum entropy problem has been adapted from that presented 
by \citet{Skilling84} which iteratively adjusts the image by a multidirection search in the 
image space;  in particular, this algorithm was found to be much more efficient than standard 
conjugate gradient techniques \citep[see, e.g.,][for more details]{Brown91}.  
The relative weights attributed to the various SH modes can be imposed, e.g., for purposedly  
producing antisymmetric or symmetric field topologies with respect to the centre of the star 
\citep[by favouring odd or even SH modes,][]{Donati07, Donati08}.  

The local synthetic photospheric Stokes $I$ and $V$ line profiles emerging from 
a given grid cell (noted \Ip\ and \Vp) are modelled using the following equations:  
\begin{eqnarray}
\Ip  & = &  b \left( \psi \Ipm + (1-\psi) \Ipq \right)  \nonumber \\
\Vp  & = &  b \, \psi \Vpm  
\label{eq:mod1}
\end{eqnarray}
where \Ipm\ and \Ipq\ are the local Stokes $I$ photospheric profiles corresponding to the magnetic 
and non-magnetic areas, \Vpm\ the local Stokes $V$ photospheric profile corresponding to the 
magnetic areas, $b$ ($0 < b \leq 1$) the local brightness relative to the quiet photosphere and $\psi$ 
($0 \leq \psi \leq 1$) the relative proportion of magnetic areas within the grid cell (called filling factor).  
For simplicity, we further assume that \Ipm\ and \Ipq\ differ by no more than magnetic effects, i.e., 
that \Ipq\ equals \Ipm\ taken at ${\bf B=0}$.  

Similarly, we describe the Stokes $I$ and $V$ \caii\ emission profiles (noted \Ie\ and \Ve) with 
equations:  
\begin{eqnarray}
\Ie & = &  \Ik + e \Ia  =  \psi (\Ikm + e \Iam) + (1-\psi) (\Ikq + e \Iaq)  \nonumber \\
\Ve & = &  \Vk + e \Va  =  \psi (\Vkm + e \Vam) 
\label{eq:mod2}
\end{eqnarray}
where \Ik\ and \Vk\ on the one hand, and \Ia\ and \Va\ on the other hand, are the respective 
contributions to $I$ and $V$ profiles from the quiet chromosphere and the accretion regions, and $e$ 
($e>0$) the local excess emission from accretion regions or equivalently the fraction of the 
grid cell occupied by accretion regions;  the 
contributions from magnetic regions to \Ik\ and \Ia\ (respectively \Vk\ and \Va) are denoted 
\Ikm\ and \Iam\ (respectively \Vkm\ and \Vam) while the contributions from non-magnetic regions 
are denoted \Ikq\ and \Iaq.  
For simplicity, we further assume that \Ia\ and \Ik\ (and similarly, \Iam\ and \Ikm, \Iaq\ and \Ikq, 
\Va\ and \Vk, \Vam\ and \Vkm) are homothetic, with a scaling factor denoted $\epsilon$ ($\epsilon>1$);  
we also suppose that \Ikm\ and \Ikq\ (and thus \Iam\ and \Iaq) differ by no more than magnetic effects 
(just as \Ipm\ and \Ipq).  The filling factor $\psi$ is finally assumed to be the same for the photospheric, 
chromospheric and accretion profiles, and for all grid cells.  
Local profiles from all grid cells are then integrated taking into account the cell visibility, 
projected area, limb-angle and radial velocity.  

To describe the local profiles (i.e., \Ipm, \Ipq, \Vpm, \Ikm, \Ikq, \Vkm), we use 
Unno-Rachkovsky's equations known to provide a good description of Stokes $I$ and $V$ profiles 
(including magneto-optical effects) in the presence of both weak and strong magnetic fields 
\citep[e.g.,][Sec.~9.8]{Landi04}  
despite their being based on the assumption of a simple Milne-Eddington atmospheric model.  
For photospheric LSD profiles (i.e., \Ipm, \Ipq, \Vpm), we set the central wavelength, Doppler width 
and Land\'e factor of our equivalent line to 640~nm, 2~\kms\ and 1.2 respectively and adjust the 
average line equivalent width to the observed value (with veiling removed from all profiles);  
for \caii\ chromospheric profiles (i.e., \Ikm, \Ikq, \Vkm), the central wavelength, Doppler 
width and Land\'e factor are 
set to 850~nm, 6.7~\kms\ and 1 respectively, the equivalent width being set to slightly below the 
minimum \caii\ emission in our observations.  

By comparing synthetic and observed profiles, we derive the spatial distributions of the local 
photospheric brightness $b$, of the \caii\ excess emission $e$ due to accretion spots, and of 
the magnetic vector $\bf B$ over the stellar surface.  As a side product, we also obtain estimates 
of several other model parameters (by selecting the image with minimum information content at a 
given fit accuracy), and in particular of the line rotational broadening \vsini, of the RV 
\vrad, of the filling factor $\psi$ and of the accretion profile scaling factor $\epsilon$.  
Simulations demonstrate that the code can reliably reconstruct magnetic topologies, brightness 
images and accretion maps;  in particular, it is found to be very efficient at recovering 
the poloidal and toroidal field components, a very useful diagnostic when studying large-scale 
magnetic topologies produced by dynamo processes at the surface of cool stars.  
Surface differential rotation patterns shearing the brightness and magnetic images can also 
be reliably retrieved, as demonstrated in its most recent application \citep{Donati10}.  

\subsection{Application to AA~Tau}

The imaging model described above assumes that the observed profile variations are mainly due to 
rotational modulation, and possibly to surface differential rotation as well when the star is 
observed for at least several rotation cycles;  all other sources of profile variability (and 
in particular intrinsic variability like flaring or short-term high-accretion episodes) cannot 
be properly reproduced and thus contribute as noise into the modelling process, degrading the 
imaging performance and potentially even drowning all relevant information.  

Filtering out significant intrinsic variability from the observed profiles of AA~Tau is thus 
worthwhile to optimise the behaviour and convergence of the imaging code, leading us to discard our 
2008 December data and to concentrate only on those collected in 2009 January and during our 
2007/2008 run.  For the latter 2 sets, we further need to suppress veiling, e.g., by scaling all 
LSD Stokes $I$ and $V$ photospheric profiles to ensure that unpolarized lines have the same 
equivalent widths;  we also need to retain rotational modulation only in the variations 
of \caii\ IRT profiles, e.g., by fitting them with a sine+cosine wave and scaling the 
corresponding Stokes $I$ and $V$ profiles to ensure that unpolarized lines match the 
fitted equivalent widths (see Fig.~\ref{fig:ew}).  
While obviously no more than approximate, this procedure has the advantage of being very 
straightforward yet reasonably efficient, and proved successful in the case of the cTTS 
V2247~Oph \citep{Donati10}.  For both runs, rotational modulation in \caii\ emission 
is found to be of order 30\% peak-to-peak.  

\begin{figure*}
\center{\hbox{
\includegraphics[scale=0.35,angle=-90]{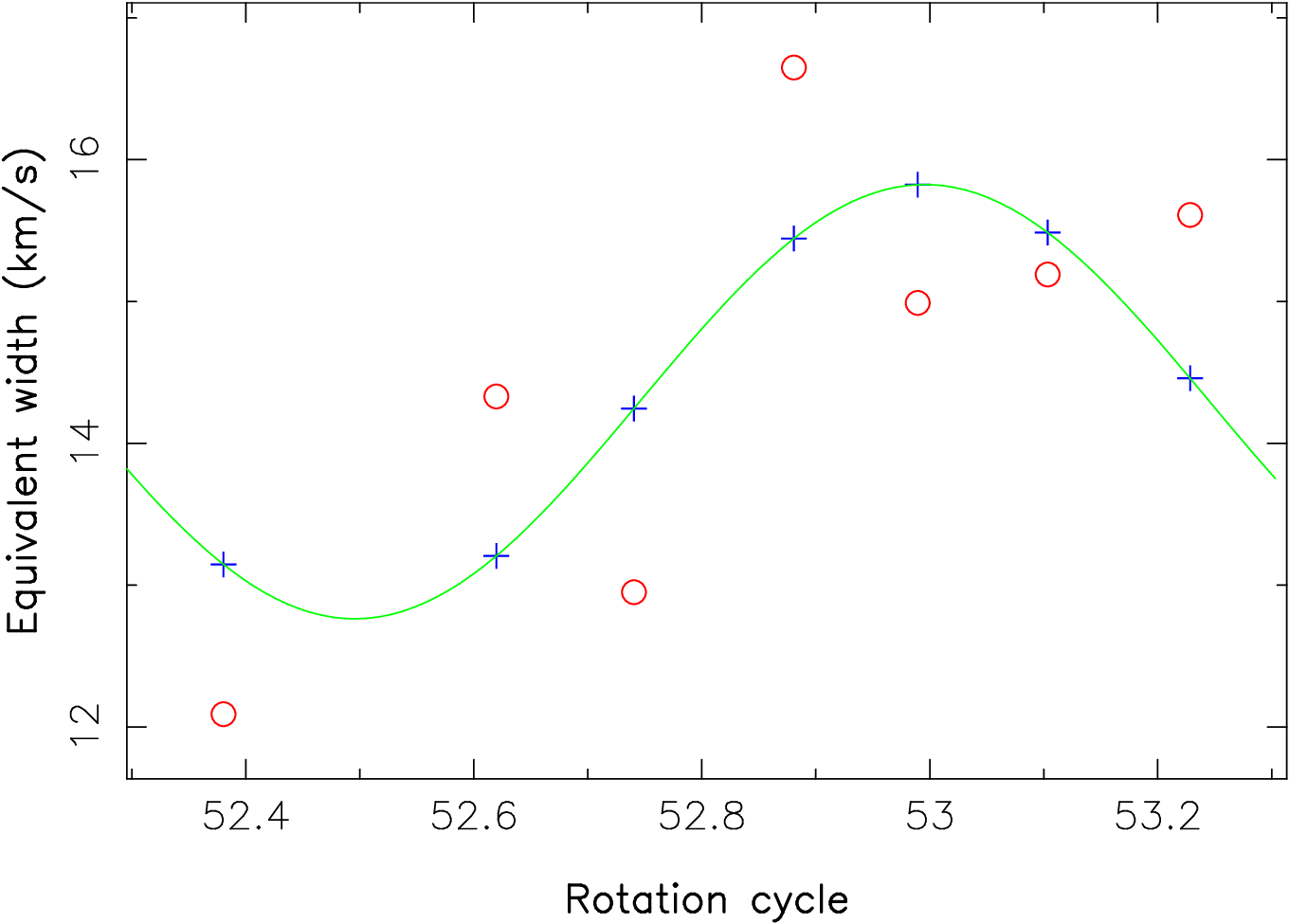}\hspace{5mm}
\includegraphics[scale=0.35,angle=-90]{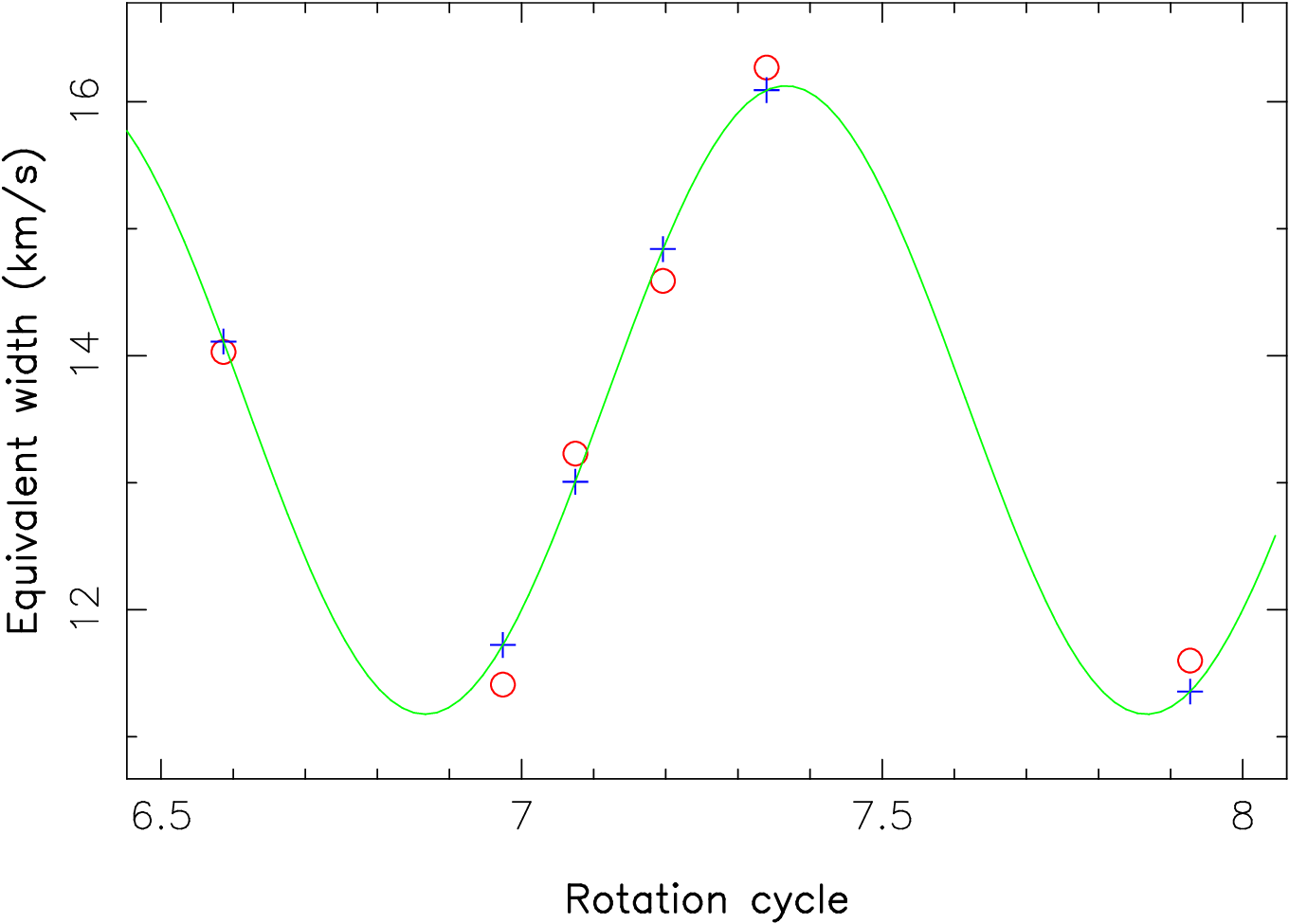}}}
\caption[]{Measured (red open circles) and fitted (blue pluses) equivalent widths of
the \caii\ IRT LSD profiles throughout our 2009 January (left) and 2007/2008 (right) 
observing runs.  The model wave (green line) providing the best (sine+cosine) fit to 
the data (with a period of 8.22~d) presumably traces rotational modulation, while the 
deviation from the fit illustrates the strength of intrinsic variability.  }
\label{fig:ew}
\end{figure*}

Given the moderate \vsini\ of AA~Tau, we limit the SH expansions describing the 
field at $\ell=5$.  
The reconstructed magnetic, brightness and accretion maps\footnote{Reconstructed 
maps are only shown down to latitudes of $-30$\degr;  being both strongly limb-darkened 
and visible for only a short amount of time, features at lower latitudes contribute little 
to the observed profiles and are therefore basically out of reach of imaging techniques, 
especially when phase coverage is moderate. }
of AA~Tau are shown in Fig.~\ref{fig:map} 
for both epochs, with corresponding fits to the data shown in 
Fig.~\ref{fig:fit}.  The overall fits are good, reproducing the data down to the 
noise level starting from initial reduced chi-squares of 18 and 5.5 for the 2009 
January and 2007/2008 data sets respectively.  Observations at both epochs (covering 
roughly a single rotation cycle) do not allow an estimate of surface differential rotation.  
Optimal fits are obtained for $\vsini=11.5\pm0.5$~\kms\ and 
$\vrad=17.2\pm0.1$~\kms, in good agreement with previous estimates 
\citep[e.g.,][]{Bouvier03, Bouvier07b};  \caii\ IRT lines are found to be slightly 
redshifted (by about 1~\kms) with respect to photospheric lines, as usual for cTTSs 
\citep[e.g.,][]{Donati07, Donati08}.  We also find that filling factors of about 
$\psi\simeq0.3\pm0.1$ provide the best fit to the far wings of the observed Stokes 
$V$ profiles (see Fig.~\ref{fig:fit}).  

The emission profile scaling factor $\epsilon$, 
describing the emission enhancement of accretion regions over the quiet chromosphere, 
is set to $\epsilon=10$.  This value is however somewhat arbitrary;  fitting the observed 
strengths of emission profiles fixes $\epsilon \sum e$ rather than $\epsilon$ alone, 
with $e$ describing how the excess emission from accretion regions varies over the surface 
of the star (see Eq~\ref{eq:mod2}) and $\sum e$ thus estimating the fractional area of 
accretion footprints.  
We nevertheless think that our choice of $\epsilon$ is realistic (and accurate within a 
typical factor of a few) given that the fractional areas of accretion footprints it yields 
(a few \%, see below) are grossly compatible with published estimates 
\citep[e.g.,][]{Valenti04}.  

\begin{figure*}
\hbox{\includegraphics[scale=0.72]{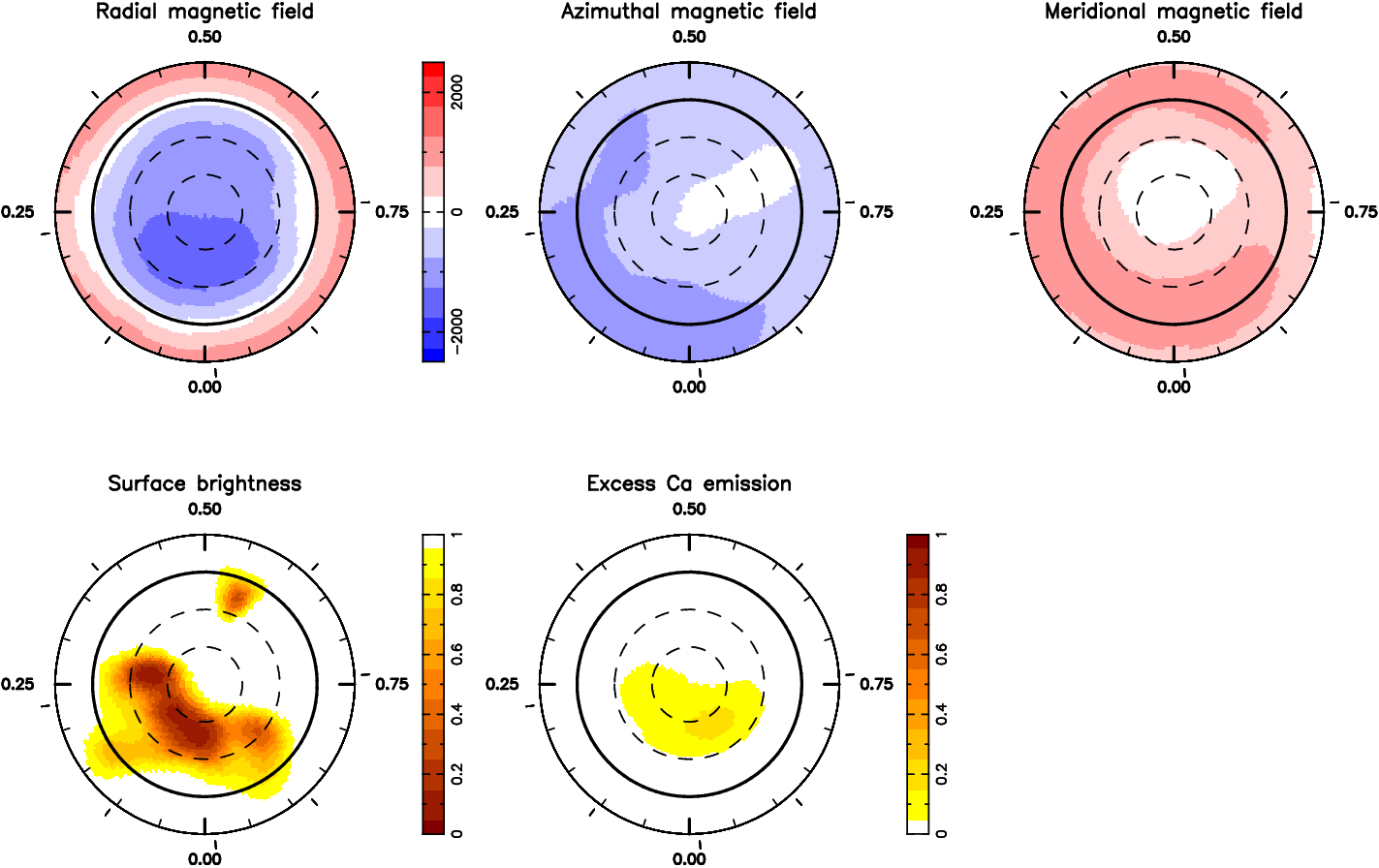}}
\vspace{8mm}
\hbox{\includegraphics[scale=0.72]{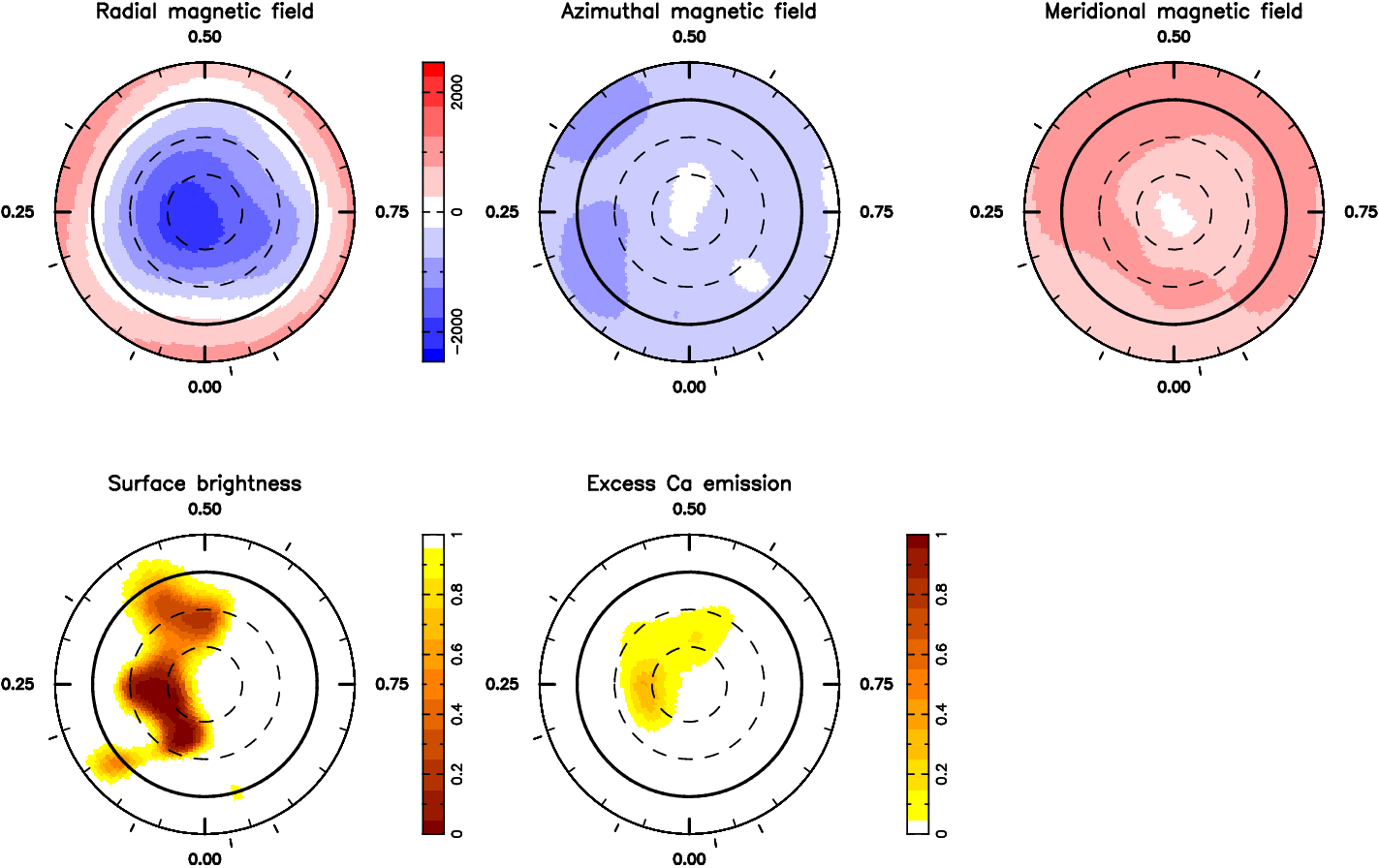}} 
\caption[]{Maps of the radial, azimuthal and meridional components of the magnetic field $\bf B$ 
(first and third rows, left to right panels respectively), photospheric brightness $b$ and excess 
\caii\ IRT emission $e$ (second and fourth rows, first and second panels respectively) at the 
surface of AA~Tau, during our 2009 January (two upper rows) and 2007/2008 (two lower rows) runs.  
Magnetic fluxes are labelled in G.  In all panels, the star is shown in flattened polar 
projection down to latitudes of $-30\degr$, with the equator depicted as a bold circle and
parallels as dashed circles.  Radial ticks around each plot indicate phases of observations. } 
\label{fig:map}
\end{figure*}

\begin{figure*}
\center{\hbox{
\hspace{-2mm}
\includegraphics[scale=0.57,angle=-90]{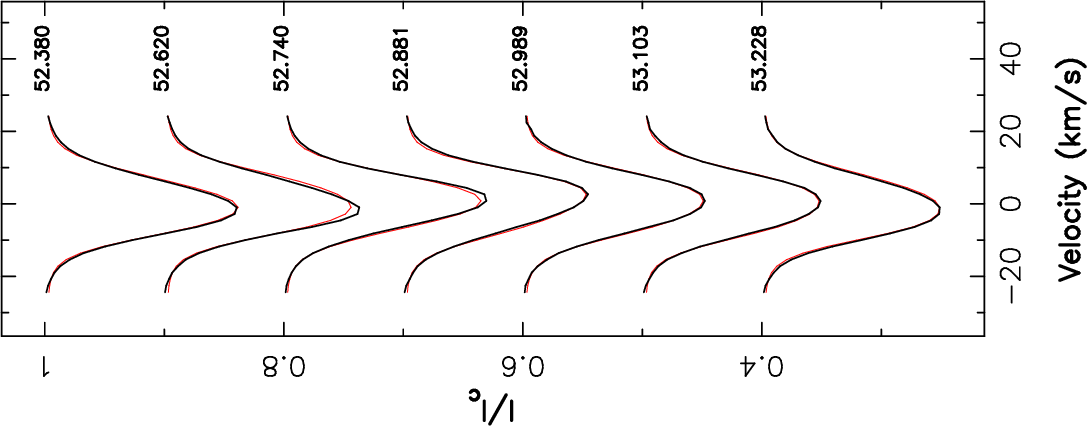}\hspace{3mm}
\includegraphics[scale=0.57,angle=-90]{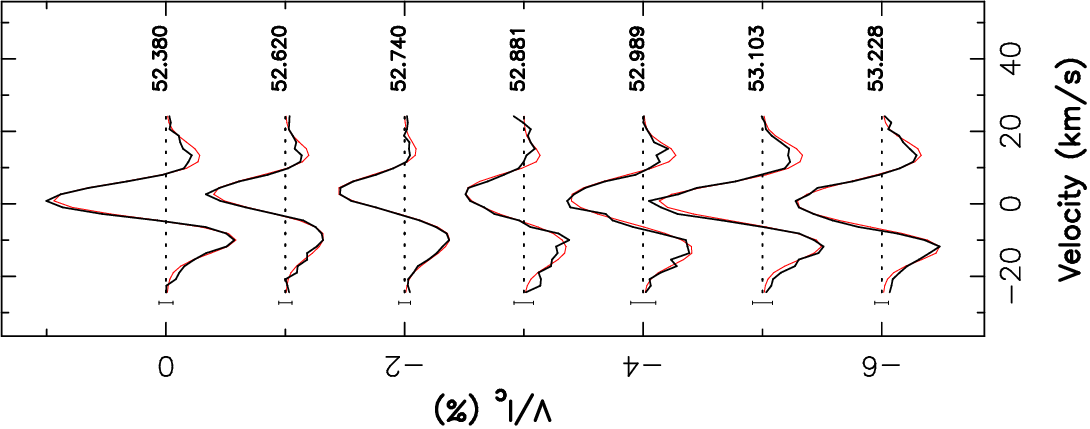}\hspace{3mm}
\includegraphics[scale=0.57,angle=-90]{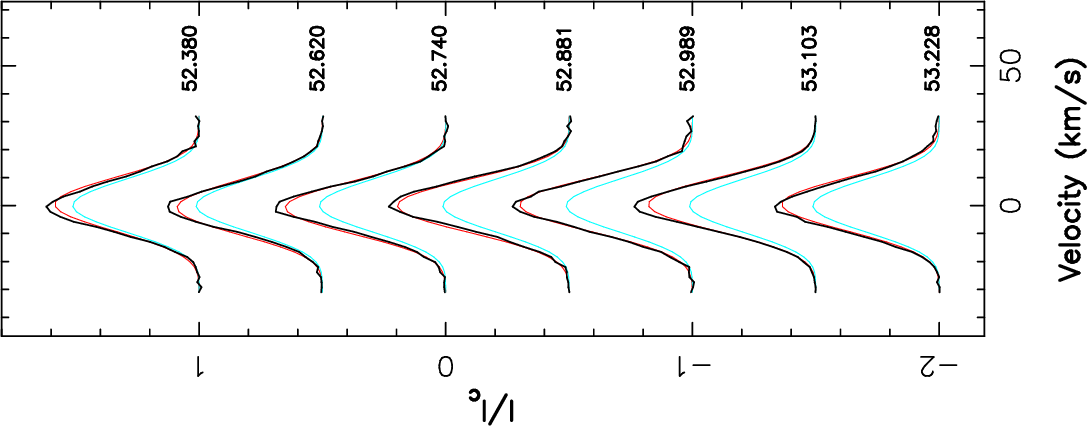}\hspace{3mm}
\includegraphics[scale=0.57,angle=-90]{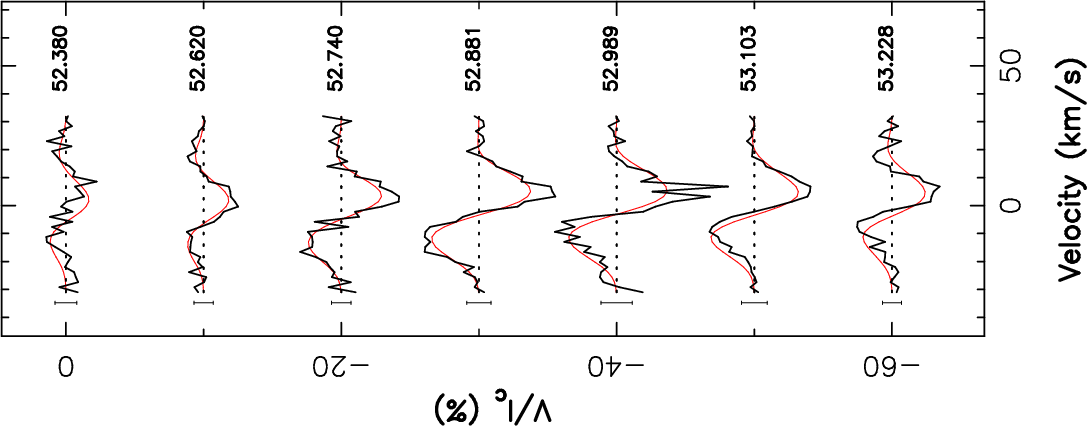}}} 
\vspace{6mm}
\center{\hbox{
\hspace{-2mm}
\includegraphics[scale=0.57,angle=-90]{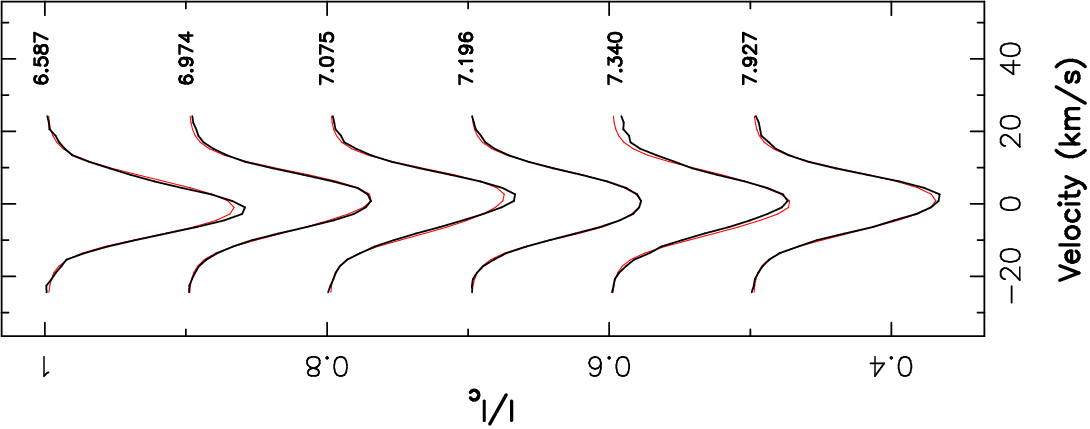}\hspace{3mm}
\includegraphics[scale=0.57,angle=-90]{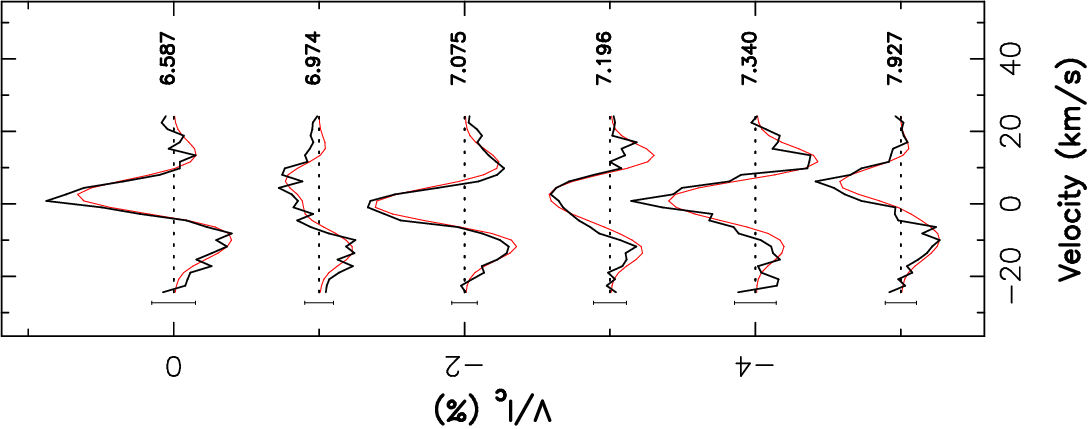}\hspace{3mm}
\includegraphics[scale=0.57,angle=-90]{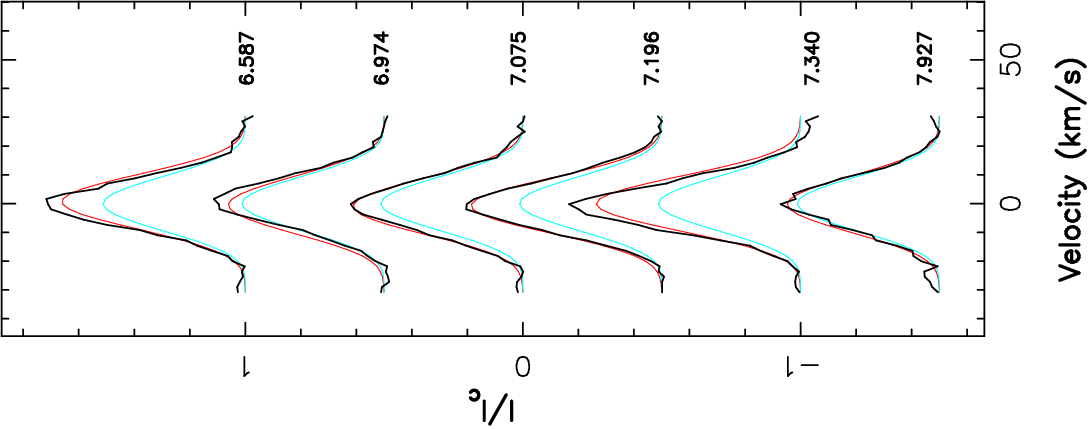}\hspace{3mm}
\includegraphics[scale=0.57,angle=-90]{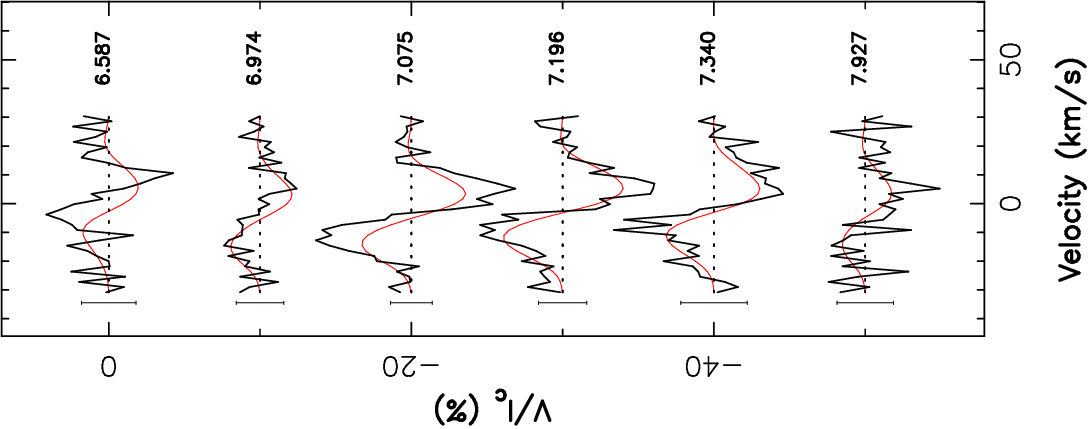}}} 
\caption[]{Maximum-entropy fit (thin red line) to the observed (thick black line) Stokes $I$ and 
Stokes $V$ LSD photospheric profiles (first two panels) and \caii\ IRT profiles (last two panels) 
of AA~Tau, for our 2009 January (upper row) and 2007/2008 (lower row) data sets.  
The light-blue curve in the third panels shows the (constant) contribution of the quiet 
chromosphere to the Stokes $I$ \caii\ profiles.
Rotation cycles and 3$\sigma$ error bars (for Stokes $V$ profiles) are also shown next to each 
profile. } 
\label{fig:fit}
\end{figure*}

The reconstructed large-scale magnetic topologies are similar at both epochs and essentially 
consist in a $\simeq2$~kG dipole inclined at about 20\degr\ to the rotation axis, 
concentrating about $90-95$\% of the poloidal field energy.  This is about 30\% weaker than 
predicted from the \hei\ alone (indicating a polar field of about 3~kG, see above), likely 
reflecting the uncertainty in the modelling of the \caii\ profiles.  We thus conservatively 
conclude that the dipole component of AA~Tau is in the range $2-3$~kG.  These estimates are
grossly compatible with previous results from dipolar fits to \hei\ spectropolarimetric data 
\citep{Valenti04}.  The quadrupole/octupole components are significantly weaker, by typically 
a factor of 5 to 10 depending on the relative weights given to 
the different SH modes (see Sec.~\ref{sec:mod}).  
We also find that the field includes a significant toroidal component in the shape of an equatorial 
ring of negative (i.e., clockwise) azimuthal field ($0.5-1$~kG), totalling about $15-20$\% 
of the magnetic energy at the surface of the star.  As already mentioned in Sec.~\ref{sec:var}, 
this toroidal component can be straightforwardly traced back to the average shape of LSD Stokes $V$ 
profiles of photospheric lines (mostly symmetric about the line centre, see Fig.~\ref{fig:pol}) and 
can thus be considered as reliable.  

The main difference between the two reconstructed magnetic images is an apparent shift of about 
0.25 cycle between the phases towards which the dipoles are tilted (phase 0 in 2009 January and 
phase 0.25 during the 2007/2008 run).  Since eclipse phases (respectively equal to 0.05 and 0.35, 
see Sec.~\ref{sec:var}) are also shifted by about the same amount, we can conclude that this shift 
essentially reflects the uncertainty on the rotation period;  a slightly smaller period (of 
$\simeq8.18$~d instead of 8.22~d, still compatible with the quoted uncertainty of 0.03~d of 
\citealt{Bouvier07b}) would bring the two images (and photometric data sets) in phase
with one another.  
Given the moderate phase coverage and the level of intrinsic variability observed on AA~Tau (even 
in Stokes $V$ profiles of photospheric lines, see Fig.~\ref{fig:var}), it is difficult to ascertain 
whether the remaining differences in the reconstructed large-scale fields at both epochs are real.  

The surface brightness maps we reconstruct are also grossly similar once accounting for the phase 
shift between both epochs.  They mostly feature a large spotted area centred at latitude 
$\simeq50$\degr, covering about 10\% of the full surface and spreading over about half the full phase 
range.  This dark area accounts in particular for the observed Stokes $I$ profile asymmetries (see 
Fig.~\ref{fig:fit} left panels) and for the corresponding RV variations.  
The location of this spot agrees well with the epochs of eclipse maxima, as expected from 
RV maxima and minima occurring respectively $\simeq$0.25 rotation cycle before and after eclipse 
maxima (see Fig.~\ref{fig:rv}).  

Given the depth of eclipses (ranging from 0.5 to 1.5~mag), it is fairly clear that this 
dark region cannot fully account by itself for the observed photometric variability;  
color and photopolarimetric variations further demonstrate that circumstellar extinction is the 
main cause of the observed brightness changes \citep[][]{Bouvier03, Menard03}.  
Similarly, one can wonder whether the reconstructed dark region is not spurious, i.e., whether 
the observed line profile variations are not the direct consequence of eclipses.  
This is unlikely for at least two reasons.  
Firstly, the varying asymmetries in the observed Stokes $I$ photospheric profiles and the 
corresponding RV fluctuations are known to occur at all times, even when eclipses are missing 
\citep[e.g., around JD~2,451,522,][]{Bouvier03};  our own data also demonstrate that RV 
variations are grossly stable and weakly affected by changes in the accretion rate and veiling 
strength (see Fig.~\ref{fig:rv} and bottom right panel of Fig.~\ref{fig:var}).  
Secondly, similar (though smaller) dark regions, more or less spatially coincident with magnetic  
poles, are observed on similar cTTSs \citep[e.g., BP~Tau,][]{Donati08}, even when not eclipsed by a 
warped inner disc (as AA~Tau).  
We thus conclude that the cool region reconstructed at the surface of AA~Tau is most likely real, 
at least its darkest parts found to grossly overlap with magnetic poles;  it is however possible 
that some of it (e.g., the trailing extension towards phase 0.45 in the 2007/2008 image, see 
lower panels of Fig.~\ref{fig:map}) is spurious and caused by minor line profile changes 
directly caused by eclipse episodes.  

\begin{figure}
\includegraphics[scale=0.35,angle=-90]{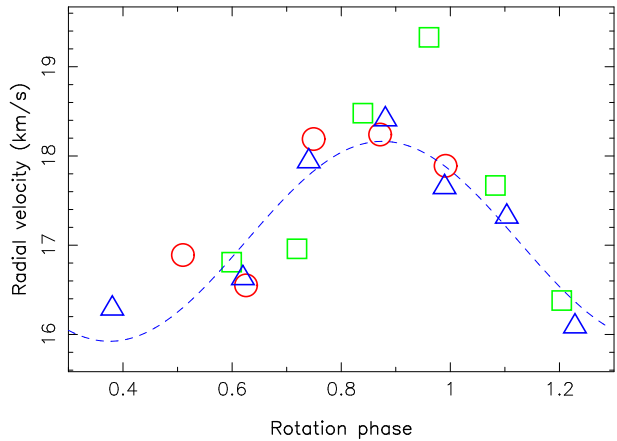}
\caption[]{RV variations of photospheric lines of AA~Tau in 2008 December and 2009 January.  
Data collected within cycles $48.3-49.3$, $49.3-50.3$ and $52.3-53.3$ are respectively plotted 
as red circles, green squares and blue triangles, with a sine/cosine fit to the blue triangles 
shown (dashed line) to outline rotational modulation. } 
\label{fig:rv}
\end{figure}

The accretion maps we derive at both epochs include one high-latitude region covering about 
2\% of the total stellar surface (when assuming $\epsilon=10$, see above) and shifted again by 
about 0.25~cycle between both runs.  This accretion region is crescent-shaped, i.e., elongated 
along parallels and spanning a phase range of about 0.2.  The brightest spot within this accretion 
region is leading, crossing the line-of-sight slightly before epochs of eclipse maxima (by about 
0.1~cycle);  it is also roughly coincident with the visible magnetic pole.  
These crescent-shaped accretion spots are very similar to those predicted in numerical simulation 
of magnetospheric accretion to slightly inclined magnetic dipoles \citep[e.g.,][]{Romanova04b}.

\section{Summary and discussion}
\label{sec:dis}

This paper presents the first results of the MaPP project aimed at studying the impact of magnetic 
fields on star and planet formation.  In particular, we report the detection of Zeeman signatures 
from the prototypical cTTS AA~Tau at two different epochs (2008 December / 2009 January and 
2007 December / 2008 January) and using the ESPaDOnS and NARVAL 
spectropolarimeters on CFHT and TBL respectively.  From phase-resolved data sets, we successfully 
derive maps of the magnetic field, of the surface brightness and of the accretion-powered emission on 
AA~Tau.  We briefly summarize below the main results and discuss their implications for our 
understanding of magnetized stellar formation.  

The large-scale magnetic field we reconstruct consists mostly of a $2-3$~kG dipole slightly 
tilted (by about 20\degr) to the rotation axis, with higher SH modes of the poloidal 
component being at least 5$\times$ smaller.  This is different from BP~Tau, the only other cTTS 
of similar mass magnetically imaged to date, where the dipole and octupole poloidal components had 
comparable strengths \citep{Donati08}.  This is even more radically different from those of both 
lower- and higher-mass cTTSs (namely V2247~Oph and MT~Ori) for which the large-scale poloidal 
field is much weaker and more complex \citep[][Skelly et al, 2010, in preparation]{Donati10};  
similar conclusions were reached for the high-mass cTTSs CV~Cha and CR~Cha \citep{Hussain09}.  
This makes AA~Tau grossly similar to mid-M dwarfs \citep[whose large-scale fields are also 
found to be strong, mostly poloidal and axisymmetric,][]{Morin08b}, while higher- and 
lower-mass cTTSs are closer to early- and late-M dwarfs respectively \citep[featuring weak and 
complex large scale fields in average,][]{Donati08c, Morin10}.  
This strengthens the idea that magnetic fields of cTTSs are produced through dynamo processes 
like those of M dwarfs \citep{Donati09}.  The origin of the differences between the large-scale 
topologies of AA~Tau and BP~Tau 
(showing very different dipole to octupole intensity ratios, $>5$ and $\simeq$1 respectively) 
are unclear, both stars featuring very similar masses, ages and 
rotation rates;  we suggest that they may reflect different magnetic states in a long-term 
magnetic cycle.  

AA~Tau also includes a small but significant toroidal component totalling about $15-20$\% of 
the reconstructed magnetic energy, i.e., about twice as much as that of BP~Tau.  This is however 
different from mid-M dwarfs for which the toroidal component rarely exceeds a few \% of the overall 
magnetic energy \citep{Morin08b}.  We speculate that this toroidal component relates to the 
accretion flow, although the link between both is not clear yet;  one option is that accretion 
is slowing down the surface of the cTTS through star/disc magnetic coupling (see below), producing 
a sub-surface shear, an additional interface dynamo (on top of the underlying dominant distributed 
dynamo operating within the bulk of the star) and the associated toroidal component.  
Regular magnetic monitoring of cTTSs is required to investigate this idea further.  

Using the \hei\ $D_3$, \caii\ IRT and H$\alpha$ emission fluxes (with stronger weight on the 
first and more reliable proxy) and the empirical correlation of \citet{Fang09}, we derive that the 
average logarithmic accretion rate at the surface of AA~Tau (in \mspy) is $-9.2\pm0.3$;  at magnetic 
maximum (i.e., phase 0.0), the logarithmic surface mass-accretion rate is found to vary by about an order 
of magnitude (from $-9.4$ up to $-8.5$) between rotation cycles, while it is roughly constant 
(at about $-9.6$) at magnetic minimum (i.e., phase 0.5).   
Given the estimated strength of the dipole field of AA~Tau ($2-3$~kG at the pole, i.e.\ $1.0-1.5$~kG 
at the equator, see above) and assuming an average logarithmic mass-accretion rate of $-9.2$, we obtain 
that the radius $\rmag$ at which the inner disc of AA~Tau is truncated should be equal to 
$\rmag\simeq15$~\rstar\ or equivalently that $\rmag/\rcor\simeq2$ \citep[using the theoretical 
estimates of][with $B_\star\simeq1.2$~kG]{Bessolaz08}.  

This result is in conflict with the conditions under which accretion can proceed, requiring 
$\rmag\leq\rcor$ to ensure that the effective gravity (i.e., the sum of gravitational and centrifugal 
forces) along accretion funnels is pointing towards the star;  $\rmag>\rcor$ implies indeed 
that effective gravity points outwards in the inner disc regions, and thus that the magnetic star 
is in the propeller regime \citep[e.g.,][]{Romanova04} and cannot accrete material from the disc, 
in contradiction with observations.  In practice, the situation is likely more complex than this 
simplistic picture;  numerical simulations of star/disc magnetic coupling in the propeller regime 
(and with $\rmag\gtrsim\rcor$) show for instance that a small fraction of the disc material can 
succeed at finding its way into the 
closed magnetosphere and onto the stellar surface even though most of it is expelled outwards 
\citep[e.g.,][]{Romanova05, Ustyugova06}.  
In this context, the mass-accretion rate at the surface of the star (i.e., the one that 
we estimate from emission proxies) is only a lower limit to the mass-accretion 
rate in the inner disc regions.  

The logarithmic accretion rate in the disc needed to enforce $\rmag\simeq\rcor$ is ranging from 
$-8.3$ to $-8.0$ (depending on the assumed value of $B_\star$), i.e., $1.6-3.2$ times larger than 
the maximum accretion rate observed at the surface of the star and $8-16$ times larger than the average 
estimated surface accretion rate.  
Interestingly enough, logarithmic accretion rates at the surface of the star derived from the 
strongest \hei\ emission levels typically reported in the literature \citep[equal to about 
75~\kms\ or 0.15~nm, e.g.,][]{Bouvier03, Bouvier07b} and tracing the highest-accretion 
phases of AA~Tau, are equal to $-8.2$, i.e., compatible with the mass-accretion rate in the disc 
that we independently derive from matching \rmag\ and \rcor.  
It confirms at least that the upper limit in the surface accretion rate that our study predicts is 
grossly compatible with published observations;  
it also suggests that AA~Tau is in a state where most of the material accreting through the disc is expelled 
outwards (the propeller regime) and only a small amount (up to 40\% in 2008 December, $\simeq10$\% in 
average, $\simeq5$\% in 2009 January) accreted towards the star.  
Our observations are in this respect qualitatively similar to results of numerical 
simulations of cTTSs in propeller regime \citep[e.g.,][]{Romanova05, Ustyugova06}.  

The strong variability that mass accretion at the surface of AA~Tau is subject to is another 
argument favouring our interpretation.  It suggests in particular that the 
accretion variability observed for AA~Tau mostly relates to the variable efficiency at which the 
disc material succeeds at entering the closed magnetosphere in propeller regime, rather than 
to an intrinsic variability of the accretion rate within the inner disc.  
This is also in qualitative agreement with results of 
numerical modelling of the propeller regime \citep[e.g.,][]{Romanova05} where the accretion 
flow from the inner disc to the surface of the star never reaches steady state and remains 
intrinsically variable.  

The systematic lag of photometric eclipses with respect to magnetic/accretion maxima (see 
Sec.~\ref{sec:var}) is another independent element favouring our schematic model.  
Previous observations of AA~Tau with the optical and UV monitors of XMM-Newton show a behaviour 
similar to that reported here, with UV maximum (presumably coinciding with magnetic and accretion 
maxima) occurring well before the optical eclipse \citep{Grosso07}, suggesting that this is 
a regular phenomenon.  
We propose that this time delay between magnetic poles and eclipse times indicates that field 
lines connecting AA~Tau to its inner accretion disc are twisted as a result of their different 
rotation rates, with leading funnel footpoints at the surface of AA~Tau and a trailing accretion 
warp (producing the lagging variable eclipse) at $\rmag\gtrsim\rcor$.  

Our study brings fresh evidence that, thanks to its strong dipole field component, AA~Tau is 
still mainly in propeller regime, with most of the material in the inner regions of the accretion 
disc being expelled outwards and only a small fraction accreted towards the star.  It suggests 
in particular that spinning down of cTTSs through star/disc magnetic coupling can potentially 
still be efficient at ages of about $1-2$~Myr.  AA~Tau thus appears as an optimal laboratory for 
studying in more details the spinning down of cTTSs in propeller regime 
and for testing predictions of numerical simulations on this issue.  
With similarly detailed analyses on a dozen of cTTSs, MaPP should soon bring important 
new material for our understanding of magnetospheric processes and their impact on the 
angular momentum evolution of forming Suns.

\section*{Acknowledgements}

This paper is based on observations obtained at the Canada-France-Hawaii
Telescope (CFHT) and at the T\'elescope Bernard Lyot (TBL).  CFHT is operated by the National
Research Council of Canada, the Institut National des Sciences de l'Univers of the Centre
National de la Recherche Scientifique of France (INSU/CNRS) and the University of Hawaii;  TBL
is operated by INSU/CNRS. 

We thank the CFHT/QSO and TBL staff for their efficiency at collecting data, as well as the 
referee,  John Landstreet, for valuable comments that improved the paper.  

The ``Magnetic Protostars and Planets'' (MaPP) project is supported by the 
funding agencies of CFHT and TBL (through the allocation of telescope time) 
and by CNRS/INSU in particular, as well as by the French ``Agence Nationale 
pour la Recherche'' (ANR).
SGG acknowledges support by the Science and Technology
Facilities Council [grant number ST/G006261/1].  

\bibliography{aatau}

\begin{thebibliography}{}

\bibitem[\protect\citeauthoryear{{Bessell}, {Castelli} \& {Plez}}{{Bessell}
  et~al.}{1998}]{Bessell98}
{Bessell} M.~S.,  {Castelli} F.,    {Plez} B.,  1998, \aap, 333, 231

\bibitem[\protect\citeauthoryear{{Bessolaz}, {Zanni}, {Ferreira}, {Keppens} \&
  {Bouvier}}{{Bessolaz} et~al.}{2008}]{Bessolaz08}
{Bessolaz} N.,  {Zanni} C.,  {Ferreira} J.,  {Keppens} R.,    {Bouvier} J.,
  2008, \aap, 478, 155

\bibitem[\protect\citeauthoryear{{Bouvier}, {Alencar}, {Boutelier}, {Dougados},
  {Balog}, {Grankin}, {Hodgkin}, {Ibrahimov}, {Kun}, {Magakian} \&
  {Pinte}}{{Bouvier} et~al.}{2007b}]{Bouvier07b}
{Bouvier} J.,  {Alencar} S.~H.~P.,  {Boutelier} T.,  {Dougados} C.,  {Balog}
  Z.,  {Grankin} K.,  {Hodgkin} S.~T.,  {Ibrahimov} M.~A.,  {Kun} M.,
  {Magakian} T.~Y.,    {Pinte} C.,  2007b, \aap, 463, 1017

\bibitem[\protect\citeauthoryear{{Bouvier}, {Alencar}, {Harries}, {Johns-Krull}
  \& {Romanova}}{{Bouvier} et~al.}{2007a}]{Bouvier07}
{Bouvier} J.,  {Alencar} S.~H.~P.,  {Harries} T.~J.,  {Johns-Krull} C.~M.,
  {Romanova} M.~M.,  2007a, in {Reipurth} B.,  {Jewitt} D.,   {Keil} K.,  eds,
  Protostars and Planets V {Magnetospheric Accretion in Classical T Tauri
  Stars}.
pp 479--494

\bibitem[\protect\citeauthoryear{{Bouvier}, {Chelli}, {Allain}, {Carrasco},
  {Costero}, {Cruz-Gonzalez}, {Dougados}, {Fern{\'a}ndez}, {Mart{\'{\i}}n},
  {M{\'e}nard}, {Mennessier}, {Mujica}, {Recillas}, {Salas}, {Schmidt} \&
  {Wichmann}}{{Bouvier} et~al.}{1999}]{Bouvier99}
{Bouvier} J.,  {Chelli} A.,  {Allain} S.,  {Carrasco} L.,  {Costero} R.,
  {Cruz-Gonzalez} I.,  {Dougados} C.,  {Fern{\'a}ndez} M.,  {Mart{\'{\i}}n}
  E.~L.,  {M{\'e}nard} F.,  {Mennessier} C.,  {Mujica} R.,  {Recillas} E.,
  {Salas} L.,  {Schmidt} G.,    {Wichmann} R.,  1999, \aap, 349, 619

\bibitem[\protect\citeauthoryear{{Bouvier}, {Grankin}, {Alencar}, {Dougados},
  {Fern{\'a}ndez}, {Basri}, {Batalha}, {Guenther}, {Ibrahimov}, {Magakian},
  {Melnikov}, {Petrov}, {Rud} \& {Zapatero Osorio}}{{Bouvier}
  et~al.}{2003}]{Bouvier03}
{Bouvier} J.,  {Grankin} K.~N.,  {Alencar} S.~H.~P.,  {Dougados} C.,
  {Fern{\'a}ndez} M.,  {Basri} G.,  {Batalha} C.,  {Guenther} E.,  {Ibrahimov}
  M.~A.,  {Magakian} T.~Y.,  {Melnikov} S.~Y.,  {Petrov} P.~P.,  {Rud} M.~V.,
   {Zapatero Osorio} M.~R.,  2003, \aap, 409, 169

\bibitem[\protect\citeauthoryear{{Brown}, {Donati}, {Rees} \& {Semel}}{{Brown}
  et~al.}{1991}]{Brown91}
{Brown} S.~F.,  {Donati} J.-F.,  {Rees} D.~E.,    {Semel} M.,  1991, \aap, 250,
  463

\bibitem[\protect\citeauthoryear{{Cieza}, {Schreiber}, {Romero}, {Mora},
  {Merin}, {Swift}, {Orellana}, {Williams}, {Harvey} \& {Evans}}{{Cieza}
  et~al.}{2010}]{cieza10}
{Cieza} L.~A.,  {Schreiber} M.~R.,  {Romero} G.~A.,  {Mora} M.~D.,  {Merin} B.,
   {Swift} J.~J.,  {Orellana} M.,  {Williams} J.~P.,  {Harvey} P.~M.,
  {Evans} N.~J.,  2010, \apj, 712, 925

\bibitem[\protect\citeauthoryear{{Donati} \& {Landstreet}}{{Donati} \&
  {Landstreet}}{2009}]{Donati09}
{Donati} J.,  {Landstreet} J.~D.,  2009, \araa, 47, 333

\bibitem[\protect\citeauthoryear{{Donati}, {Skelly}, {Bouvier}, {Jardine},
  {Gregory}, {Morin}, {Hussain}, {Dougados}, {M{\'e}nard} \& {Unruh}}{{Donati}
  et~al.}{2010}]{Donati10}
{Donati} J.,  {Skelly} M.~B.,  {Bouvier} J.,  {Jardine} M.~M.,  {Gregory}
  S.~G.,  {Morin} J.,  {Hussain} G.~A.~J.,  {Dougados} C.,  {M{\'e}nard} F.,
  {Unruh} Y.,  2010, \mnras, 402, 1426

\bibitem[\protect\citeauthoryear{{Donati}}{{Donati}}{2003}]{Donati03}
{Donati} J.-F.,  2003, in {Trujillo-Bueno} J.,  {Sanchez Almeida} J.,  eds,
  Astronomical Society of the Pacific Conference Series Vol.~307 of
  Astronomical Society of the Pacific Conference Series, {ESPaDOnS: An Echelle
  SpectroPolarimetric Device for the Observation of Stars at CFHT}.
pp 41--+

\bibitem[\protect\citeauthoryear{{Donati}, {Howarth}, {Jardine}, {Petit},
  {Catala}, {Landstreet}, {Bouret}, {Alecian}, {Barnes}, {Forveille}, {Paletou}
  \& {Manset}}{{Donati} et~al.}{2006}]{Donati06b}
{Donati} J.-F.,  {Howarth} I.~D.,  {Jardine} M.~M.,  {Petit} P.,  {Catala} C.,
  {Landstreet} J.~D.,  {Bouret} J.-C.,  {Alecian} E.,  {Barnes} J.~R.,
  {Forveille} T.,  {Paletou} F.,    {Manset} N.,  2006, \mnras, 370, 629

\bibitem[\protect\citeauthoryear{{Donati}, {Jardine}, {Gregory}, {Petit},
  {Bouvier}, {Dougados}, {M{\'e}nard}, {Cameron}, {Harries}, {Jeffers} \&
  {Paletou}}{{Donati} et~al.}{2007}]{Donati07}
{Donati} J.-F.,  {Jardine} M.~M.,  {Gregory} S.~G.,  {Petit} P.,  {Bouvier} J.,
   {Dougados} C.,  {M{\'e}nard} F.,  {Cameron} A.~C.,  {Harries} T.~J.,
  {Jeffers} S.~V.,    {Paletou} F.,  2007, \mnras, 380, 1297

\bibitem[\protect\citeauthoryear{{Donati}, {Jardine}, {Gregory}, {Petit},
  {Paletou}, {Bouvier}, {Dougados}, {M{\'e}nard}, {Cameron}, {Harries},
  {Hussain}, {Unruh}, {Morin}, {Marsden}, {Manset}, {Auri{\`e}re}, {Catala} \&
  {Alecian}}{{Donati} et~al.}{2008a}]{Donati08}
{Donati} J.-F.,  {Jardine} M.~M.,  {Gregory} S.~G.,  {Petit} P.,  {Paletou} F.,
   {Bouvier} J.,  {Dougados} C.,  {M{\'e}nard} F.,  {Cameron} A.~C.,  {Harries}
  T.~J.,  {Hussain} G.~A.~J.,  {Unruh} Y.,  {Morin} J.,  {Marsden} S.~C.,
  {Manset} N.,  {Auri{\`e}re} M.,  {Catala} C.,    {Alecian} E.,  2008a, \mnras,
  386, 1234

\bibitem[\protect\citeauthoryear{{Donati}, {Morin}, {Petit}, {Delfosse},
  {Forveille}, {Auri{\`e}re}, {Cabanac}, {Dintrans}, {Fares}, {Gastine},
  {Jardine}, {Ligni{\`e}res}, {Paletou}, {Velez} \& {Th{\'e}ado}}{{Donati}
  et~al.}{2008c}]{Donati08c}
{Donati} J.-F.,  {Morin} J.,  {Petit} P.,  {Delfosse} X.,  {Forveille} T.,
  {Auri{\`e}re} M.,  {Cabanac} R.,  {Dintrans} B.,  {Fares} R.,  {Gastine} T.,
  {Jardine} M.~M.,  {Ligni{\`e}res} F.,  {Paletou} F.,  {Velez} J.~C.~R.,
  {Th{\'e}ado} S.,  2008c, \mnras, 390, 545

\bibitem[\protect\citeauthoryear{{Donati}, {Moutou}, {Far{\`e}s}, {Bohlender},
  {Catala}, {Deleuil}, {Shkolnik}, {Cameron}, {Jardine} \& {Walker}}{{Donati}
  et~al.}{2008b}]{Donati08b}
{Donati} J.-F.,  {Moutou} C.,  {Far{\`e}s} R.,  {Bohlender} D.,  {Catala} C.,
  {Deleuil} M.,  {Shkolnik} E.,  {Cameron} A.~C.,  {Jardine} M.~M.,    {Walker}
  G.~A.~H.,  2008b, \mnras, 385, 1179

\bibitem[\protect\citeauthoryear{{Donati}, {Paletou}, {Bouvier} \&
  {Ferreira}}{{Donati} et~al.}{2005}]{Donati05}
{Donati} J.-F.,  {Paletou} F.,  {Bouvier} J.,    {Ferreira} J.,  2005, \nat,
  438, 466

\bibitem[\protect\citeauthoryear{{Donati}, {Semel}, {Carter}, {Rees} \&
  {Collier Cameron}}{{Donati} et~al.}{1997}]{Donati97b}
{Donati} J.-F.,  {Semel} M.,  {Carter} B.~D.,  {Rees} D.~E.,    {Collier
  Cameron} A.,  1997, \mnras, 291, 658

\bibitem[\protect\citeauthoryear{{Fang}, {van Boekel}, {Wang}, {Carmona},
  {Sicilia-Aguilar} \& {Henning}}{{Fang} et~al.}{2009}]{Fang09}
{Fang} M.,  {van Boekel} R.,  {Wang} W.,  {Carmona} A.,  {Sicilia-Aguilar} A.,
    {Henning} T.,  2009, ArXiv e-prints

\bibitem[\protect\citeauthoryear{{Grankin}, {Melnikov}, {Bouvier}, {Herbst} \&
  {Shevchenko}}{{Grankin} et~al.}{2007}]{Grankin07}
{Grankin} K.~N.,  {Melnikov} S.~Y.,  {Bouvier} J.,  {Herbst} W.,
  {Shevchenko} V.~S.,  2007, \aap, 461, 183

\bibitem[\protect\citeauthoryear{{Gregory}, {Matt}, {Donati} \&
  {Jardine}}{{Gregory} et~al.}{2008}]{Gregory08}
{Gregory} S.~G.,  {Matt} S.~P.,  {Donati} J.-F.,    {Jardine} M.,  2008,
  \mnras, 389, 1839

\bibitem[\protect\citeauthoryear{{Grosso}, {Bouvier}, {Montmerle},
  {Fern{\'a}ndez}, {Grankin} \& {Zapatero Osorio}}{{Grosso}
  et~al.}{2007}]{Grosso07}
{Grosso} N.,  {Bouvier} J.,  {Montmerle} T.,  {Fern{\'a}ndez} M.,  {Grankin}
  K.,    {Zapatero Osorio} M.~R.,  2007, \aap, 475, 607

\bibitem[\protect\citeauthoryear{{Hussain}, {Collier Cameron}, {Jardine},
  {Dunstone}, {Velez}, {Stempels}, {Donati}, {Semel}, {Aulanier}, {Harries},
  {Bouvier}, {Dougados}, {Ferreira}, {Carter} \& {Lawson}}{{Hussain}
  et~al.}{2009}]{Hussain09}
{Hussain} G.~A.~J.,  {Collier Cameron} A.,  {Jardine} M.~M.,  {Dunstone} N.,
  {Velez} J.~R.,  {Stempels} H.~C.,  {Donati} J.-F.,  {Semel} M.,  {Aulanier}
  G.,  {Harries} T.,  {Bouvier} J.,  {Dougados} C.,  {Ferreira} J.,  {Carter}
  B.~D.,    {Lawson} W.~A.,  2009, \mnras, pp 997--+

\bibitem[\protect\citeauthoryear{{Jardine}, {Gregory} \& {Donati}}{{Jardine}
  et~al.}{2008}]{Jardine08}
{Jardine} M.~M.,  {Gregory} S.~G.,    {Donati} J.-F.,  2008, \mnras, 386, 688

\bibitem[\protect\citeauthoryear{{Johns-Krull}}{{Johns-Krull}}{2007}]{Johns07}
{Johns-Krull} C.~M.,  2007, \apj, 664, 975

\bibitem[\protect\citeauthoryear{{Kurucz}}{{Kurucz}}{1993}]{Kurucz93}
{Kurucz} R.,  1993, CDROM \#~13 (ATLAS9 atmospheric models) and \#~18 (ATLAS9
  and SYNTHE routines, spectral line database).
Smithsonian Astrophysical Observatory, Washington D.C.

\bibitem[\protect\citeauthoryear{{Landi degl'Innocenti} \& {Landolfi}}{{Landi
  degl'Innocenti} \& {Landolfi}}{2004}]{Landi04}
{Landi degl'Innocenti} E.,  {Landolfi} M.,  2004, {Polarisation in spectral
  lines}.
Dordrecht/Boston/London: Kluwer Academic Publishers

\bibitem[\protect\citeauthoryear{{M{\'e}nard}, {Bouvier}, {Dougados},
  {Mel'nikov} \& {Grankin}}{{M{\'e}nard} et~al.}{2003}]{Menard03}
{M{\'e}nard} F.,  {Bouvier} J.,  {Dougados} C.,  {Mel'nikov} S.~Y.,
  {Grankin} K.~N.,  2003, \aap, 409, 163

\bibitem[\protect\citeauthoryear{{Morin}, {Donati}, {Petit}, {Delfosse},
  {Forveille} \& {Jardine}}{{Morin} et~al.}{2010}]{Morin10}
{Morin} J.,  {Donati} J.,  {Petit} P.,  {Delfosse} X.,  {Forveille} T.,
  {Jardine} M.~M.,  2010, \mnras, pp 1077--+

\bibitem[\protect\citeauthoryear{{Morin}, {Donati}, {Petit}, {Delfosse},
  {Forveille}, {Albert}, {Auri{\`e}re}, {Cabanac}, {Dintrans}, {Fares},
  {Gastine}, {Jardine}, {Ligni{\`e}res}, {Paletou}, {Ramirez Velez} \&
  {Th{\'e}ado}}{{Morin} et~al.}{2008}]{Morin08b}
{Morin} J.,  {Donati} J.-F.,  {Petit} P.,  {Delfosse} X.,  {Forveille} T.,
  {Albert} L.,  {Auri{\`e}re} M.,  {Cabanac} R.,  {Dintrans} B.,  {Fares} R.,
  {Gastine} T.,  {Jardine} M.~M.,  {Ligni{\`e}res} F.,  {Paletou} F.,  {Ramirez
  Velez} J.~C.,    {Th{\'e}ado} S.,  2008, \mnras, 390, 567

\bibitem[\protect\citeauthoryear{{Natta}, {Testi}, {Muzerolle}, {Randich},
  {Comer{\'o}n} \& {Persi}}{{Natta} et~al.}{2004}]{Natta04}
{Natta} A.,  {Testi} L.,  {Muzerolle} J.,  {Randich} S.,  {Comer{\'o}n} F.,
  {Persi} P.,  2004, \aap, 424, 603

\bibitem[\protect\citeauthoryear{{Pojmanski}}{{Pojmanski}}{1997}]{Pojmanski97}
{Pojmanski} G.,  1997, Acta Astronomica, 47, 467

\bibitem[\protect\citeauthoryear{{Romanova}, {Ustyugova}, {Koldoba} \&
  {Lovelace}}{{Romanova} et~al.}{2004a}]{Romanova04}
{Romanova} M.~M.,  {Ustyugova} G.~V.,  {Koldoba} A.~V.,    {Lovelace} R.~V.~E.,
   2004a, \apjl, 616, L151

\bibitem[\protect\citeauthoryear{{Romanova}, {Ustyugova}, {Koldoba} \&
  {Lovelace}}{{Romanova} et~al.}{2004b}]{Romanova04b}
{Romanova} M.~M.,  {Ustyugova} G.~V.,  {Koldoba} A.~V.,    {Lovelace} R.~V.~E.,
   2004b, \apj, 610, 920

\bibitem[\protect\citeauthoryear{{Romanova}, {Ustyugova}, {Koldoba} \&
  {Lovelace}}{{Romanova} et~al.}{2005}]{Romanova05}
{Romanova} M.~M.,  {Ustyugova} G.~V.,  {Koldoba} A.~V.,    {Lovelace} R.~V.~E.,
   2005, \apjl, 635, L165

\bibitem[\protect\citeauthoryear{{Siess}, {Dufour} \& {Forestini}}{{Siess}
  et~al.}{2000}]{Siess00}
{Siess} L.,  {Dufour} E.,    {Forestini} M.,  2000, \aap, 358, 593

\bibitem[\protect\citeauthoryear{{Skilling} \& {Bryan}}{{Skilling} \&
  {Bryan}}{1984}]{Skilling84}
{Skilling} J.,  {Bryan} R.,  1984, \mnras, 211, 111

\bibitem[\protect\citeauthoryear{{Ustyugova}, {Koldoba}, {Romanova} \&
  {Lovelace}}{{Ustyugova} et~al.}{2006}]{Ustyugova06}
{Ustyugova} G.~V.,  {Koldoba} A.~V.,  {Romanova} M.~M.,    {Lovelace} R.~V.~E.,
   2006, \apj, 646, 304

\bibitem[\protect\citeauthoryear{{Valenti} \& {Johns-Krull}}{{Valenti} \&
  {Johns-Krull}}{2004}]{Valenti04}
{Valenti} J.,  {Johns-Krull} C.,  2004, \apss, 292, 619

\end{thebibliography}

\bibliographystyle{mn2e}

\end{document}